\documentclass{revtex4}%
\usepackage{amsfonts}
\usepackage{amsmath}
\usepackage{amssymb}
\usepackage{graphicx}
\usepackage{hyperref}
\usepackage{bm}%
\begin{document}
\title{Simulated annealing for generalized Skyrme models}
\author{J.-P. Longpr\'e and L. Marleau}
\email{lmarleau@phy.ulaval.ca}
\affiliation{D\'epartement de Physique, de G\'enie Physique et d'Optique, Universit\'e
Laval, Qu\'ebec, Qu\'ebec, Canada G1K 7P4}
\date{\today }

\begin{abstract}
We use a simulated annealing algorithm to find the static field configuration
with the lowest energy in a given sector of topological charge for generalized
$SU(2)$ Skyrme models. These numerical results suggest that the following
conjecture may hold: the symmetries of the soliton solutions of extended
Skyrme models are the same as for the Skyrme model. Indeed, this is verified
for two effective Lagrangians with terms of order six and order eight in
derivatives of the pion fields respectively for topological charges $B=1$ up
to $B=4$. We also evaluate the energy of these multi-skyrmions using the
rational maps ansatz. A comparison with the exact numerical results shows that
the reliability of this approximation for extended Skyrme models is almost as
good as for the pure Skyrme model. Some details regarding the implementation
of the simulated annealing algorithm in one and three spatial dimensions are provided.

\end{abstract}
\maketitle

\section{\label{sec:intro}Introduction}

The Skyrme model \cite{Skyrme1961} was originally formulated to provide a
description of baryons as topological solitons of finite energy emerging in
the framework of a nonlinear theory of weakly coupled pions fields. Nowadays,
this idea is partly supported by the $1/N_{c}$ analysis \cite{'tHooft1974,
Witten1979}, according to which the low-energy limit of QCD could be
represented by an effective theory of infinitely many mesons fields whose
derivatives appear to all-orders. Since little is known about the exact form
of such a Lagrangian, significant efforts have been made to formulate in a
simple way Skyrme-like effective Lagrangians \cite{Jackson1985, Marleau1990}.
It was then possible to improve the phenomenological predictions for the
spherically symmetric skyrmion with unit topological charge $(B=1)$
\cite{DubeMarleau1990, Jackson1991} with higher order Lagrangians, whereas
only a relative accord with experimental data was achieved with the original
Skyrme model \cite{Adkins1983}.

On the other hand, analysis based on axially symmetric ansatz
\cite{Kopeliovich1987,Bratten1988} and full numerical studies
\cite{Bratten1990, Battye19971, Battye19972} have shown that the angular
distribution of static Skyrme fields is not accurately reproduced by the
spherically symmetric hedgehog ansatz in the general context of topological
charges $B>1$. For the case of models containing higher order terms, not much
is known for these $B>1$ and the angular configuration of skyrmions needs to
be investigated in greater details. A few steps towards this has been made in
\cite{Kopeliovich1987} and more recently in \cite{Floratos2001} for a Skyrme
model extended to order six in derivatives of the pion fields assuming an
axially symmetric solution.

In this work, we use a simulated annealing algorithm to minimize the static
energy functional of order six and order eight extended Skyrme models for
baryonic numbers $B=1$ to $B=4$, following the approach in \cite{Hale2000}
where this numerical method was used to solve the Skyrme model. No specific
ansatz is used to get the minimum energy solutions. The essence of simulated
annealing (SA) relies on the analogy that can be made with a solid which is
slowly cooled down, stating that if thermal equilibrium is achieved at each
temperature during the cooling process, the solid will eventually reach its
ground state. The first application of this idea to optimization problems like
the minimization of energy functionals has been demonstrated in
\cite{Kirkpatrick1983}. We use a similar strategy where SA describes the
cooling process of our system and a Metropolis algorithm \cite{Metropolis1953}
brings it into thermal equilibrium. The main advantage of this method over
other techniques is that it only involves the energy density and there is no
need to write or solve directly a set of differential equations that become
more complex as the order of the Lagrangian increases.

The exact soliton solutions that we obtain with the SA algorithm also
constitute a valuable comparison tool to study the rational maps approximation
for Skyrme fields in the case of generalized models. This approximation works
quite well for the Skyrme model \cite{Houghton1998, Battye2002}, and one might
conjecture that a rational map based ansatz would accurately depict the
solutions of extended Skyrme models in general. Interestingly, it has been
demonstrated that adding a sixth order term in derivatives of the pion fields
to the Skyrme Lagrangian does not compromise the reliability of the rational
maps ansatz \cite{Floratos2001}. Conversely, when a pion mass term is
introduced, the symmetries of the skyrmions differ from those of the original
Skyrme model for higher topological sectors \cite{Battye2004}, making the use
of rational maps inappropriate. Thus, it is pertinent to test further this
ansatz for other extensions of the Skyrme model.

In the next section, we give a brief account on the structure of the Skyrme
model and some of its extensions. Then, in section \ref{sec:RatMaps}, we
review the rational maps ansatz and use it to write a static energy functional
for generalized Skyrme models. As an aside, we introduce a new positivity
constraint on the models proposed in \cite{DubeMarleau1990, Jackson1991} in
section \ref{sec:Positivity}. Comparison with rational maps ansatz requires a
finite order Lagrangian, which motivates our choice to restrict our analysis
to order-six and order-eight models. Section \ref{sec:SimAnn} provides a
rather detailed description of the SA algorithm we use in three spatial
dimensions, leading to exact soliton solutions, and the one we use in one
spatial dimension to minimize the radial energy functional of the skyrmion,
where the angular dependence has already been integrated out as a result of
the use of rational maps. Finally, we discuss the validity and the limits of
our SA results in section \ref{sec:Results}, where we also draw conclusions
concerning the symmetries of the skyrmions we obtain and the reliability of
the rational maps ansatz.

\section{\label{sec:Skyrme}Generalized Skyrme models}

The $SU(2)$ Skyrme model Lagrangian density for zero pion mass takes the form
\begin{equation}
\mathcal{L}_{S}=-\frac{F_{\pi}^{2}}{16}~\mathrm{Tr}L_{\mu}L^{\mu}+\frac
{1}{32e^{2}}~\mathrm{Tr}f_{\mu\nu}f^{\mu\nu}, \label{eq:Skyrme_lag}%
\end{equation}
where $L_{\mu}=U^{\dag}\partial_{\mu}U$ is a left-handed chiral current and
$f_{\mu\nu}\equiv\left[  L_{\mu},L_{\nu}\right]  $. The $SU(2)$ chiral field
$U$ is a matrix whose degrees of freedom can be parametrized in terms of the
$\sigma$ and $\bm{\pi}$ fields using $U=\frac{2}{F_{\pi}}(\sigma
+i\bm{\tau}\cdot\bm{\pi})$ with $\sigma^{2}+\bm{\pi}^{2}=\frac{F_{\pi}^{2}}%
{4}$, providing then a link with a nonlinear pion theory. The first term in
(\ref{eq:Skyrme_lag}) coincides with the nonlinear sigma model which admits
soliton solutions that are unstable with regard to a scale transformation.
Skyrme proposed to add a term of order four in derivatives of the pion fields,
the second term in (\ref{eq:Skyrme_lag}), to stabilize the solitons and
account for nucleon-nucleon interactions via pion exchange. The parameter
$F_{\pi}$ is the pion decay coupling and $e$ is a constant dimensionless
coupling. Both parameters are usually fixed using nucleon properties. Choosing
an appropriate change of variables, we rewrite from hereon the Lagrangian
density (\ref{eq:Skyrme_lag}) using units of length $(2\sqrt{2})/(eF_{\pi})$
and of energy $F_{\pi}/(2\sqrt{2}e)$ leading to
\begin{equation}
\mathcal{L}_{S}=\mathcal{L}_{1}+\frac{1}{2}~\mathcal{L}_{2}=\left(  -\frac
{1}{2}~\mathrm{Tr}L_{\mu}L^{\mu}\right)  +\frac{1}{2}\left(  \frac{1}%
{16}~\mathrm{Tr}f_{\mu\nu}f^{\mu\nu}\right)  .
\end{equation}

To obtain finite energy solutions, the field configuration $U$ must respect
the boundary condition $U\rightarrow\mathbf{1}$ at spatial infinity, stating
that this map from $R^{3}$ to $SU(2)$ goes to the trivial vacuum for
asymptotically large distances. Each mapping $U$ is characterized by an
integer topological invariant, the winding number. Skyrme associated this
conserved quantity with the baryon number
\begin{equation}
B=-\frac{1}{24\pi^{2}}\int d^{3}x~\epsilon_{ijk}\mathrm{Tr}\left(  L_{i}%
L_{j}L_{k}\right)  . \label{eq:charge}%
\end{equation}

From physical grounds, the order-four stabilizing term added by Skyrme in
(\ref{eq:Skyrme_lag}) is somewhat arbitrary as it leads to a chiral theory
which is expected to be valid only at low momenta. So, the search for an
effective Lagrangian more appropriate for the description of low-energy QCD
properties prompted naturally the inclusion of higher-order terms in
derivatives of the pion fields to the Skyrme model. But even writing the most
general higher-order Lagrangian rapidly becomes a cumbersome task as the
number of terms increases with the number of derivatives let alone finding any
solutions. In view of this difficulty, we choose to consider only a class of
tractable models defined in \cite{Marleau1990}. The reason for such a choice
will be explained in the next sections. For now, let us mention that chiral
symmetry is preserved to all-orders in derivatives of the pion fields.
Following this scheme, the most general Lagrangian $\mathcal{L}$ takes the
form
\begin{equation}
\mathcal{L}=\sum_{m=1}^{\infty}h_{m}\mathcal{L}_{m} \label{eq:L}%
\end{equation}
where
\begin{align}
\mathcal{L}_{1}  &  =-\frac{1}{2}~\mathrm{Tr}\left(  L_{\mu}L^{\mu}\right)
,\label{eq:L1}\\
\mathcal{L}_{2}  &  =\frac{1}{16}~\mathrm{Tr}\left(  f^{\mu\nu}f_{\mu\nu
}\right)  ,\label{eq:L2}\\
\mathcal{L}_{3}  &  =-\frac{1}{32}~\mathrm{Tr}\left(  f_{\mu\nu}f^{\nu\lambda
}f_{\lambda}^{~\mu}\right)  , \label{eq:L3}%
\end{align}
are respectively terms of order two, four and six in derivatives of the pion
fields. Any higher-order Lagrangians can be written in terms of $\mathcal{L}%
_{1},\mathcal{L}_{2}$ and $\mathcal{L}_{3}$ according to the recursion formula%
\begin{equation}
\mathcal{L}_{m}=-\mathcal{L}_{1}\mathcal{L}_{m-1}+\mathcal{L}_{2}%
\mathcal{L}_{m-2}-\frac{1}{3}\mathcal{L}_{3}\mathcal{L}_{m-3}\quad
\text{for}\quad m>3 \label{eq:Lm}%
\end{equation}
or using a generating function \cite{Marleau2001}. For our 3D simulations, it
is convenient to parametrize the four degrees of freedom by $\phi_{m}$ with
$m=0,1,2,3$ \ such that $\phi_{m}\phi_{m}=1.$ These fields are related to the
$\sigma$ and pion fields according to $(\phi_{0},\phi_{1},\phi_{2},\phi
_{3})=\frac{2}{F_{\pi}}(\sigma,\bm{\pi}).$ The previous Lagrangians then take
the form
\begin{align}
\mathcal{L}_{1}  &  =\partial_{\mu}\phi_{m}\partial^{\mu}\phi_{m},\\
\mathcal{L}_{2}  &  =-\frac{1}{2}~\left[  (\partial_{\mu}\phi_{m}\partial
^{\mu}\phi_{m})^{2}-(\partial_{\mu}\phi_{m}\partial_{\nu}\phi_{m})^{2}\right]
,\\
\mathcal{L}_{3}  &  =\frac{1}{2}~(\partial_{\mu}\phi_{m}\partial^{\mu}\phi
_{m})^{3}-\frac{3}{2}~(\partial_{\mu}\phi_{m}\partial^{\mu}\phi_{m}%
)(\partial_{\nu}\phi_{n}\partial_{\lambda}\phi_{n})^{2}\\
&  +(\partial_{\mu}\phi_{m}\partial_{\nu}\phi_{m})(\partial^{\nu}\phi
_{n}\partial^{\lambda}\phi_{n})(\partial_{\lambda}\phi_{l}\partial^{\mu}%
\phi_{l}).\nonumber
\end{align}

The static energy density emerging from those Lagrangians can be written in
terms of three invariants $a,b$ and $c$
\begin{align*}
\mathcal{E}_{1}  &  =a+b+c=\mathrm{Tr}\ D\\
\mathcal{E}_{2}  &  =ab+bc+ca=\frac{1}{2}\{\left(  \mathrm{Tr}\ D\right)
^{2}-\mathrm{Tr}\ D^{2}\}\\
\mathcal{E}_{3}  &  =3abc=3\det\ D
\end{align*}
and
\[
\mathcal{E}_{m}=\mathcal{E}_{1}\mathcal{E}_{m-1}-\mathcal{E}_{2}%
\mathcal{E}_{m-2}+\frac{1}{3}\mathcal{E}_{3}\mathcal{E}_{m-3}\quad
\text{for}\quad m>3.
\]
Manton has shown that these invariants have a simple geometrical
interpretation \cite{Manton1987}. They correspond to the eigenvalues of the
strain tensor $D_{ij}=\partial_{i}\phi_{m}\partial_{j}\phi_{m}$ in the theory
of elasticity, i.e. the square of the length changes of the images of any
orthonormal system in the space manifold $R^{3}$ under the conformal map $U$
onto the group manifold $SU(2)\cong S^{3}$. Accordingly, $\mathcal{E}_{1}$,
$\mathcal{E}_{2}$ and $\mathcal{E}_{3}$ may be interpreted as $\sum\left(
\text{length}\right)  ^{2}$, $\sum\left(  \text{area}\right)  ^{2}$ and
$3\left(  \text{volume}\right)  ^{2}$. The total energy density associated
with a model can then be recast as
\begin{equation}
\mathcal{E}=\sum_{m=1}^{\infty}h_{m}~\mathcal{E}_{m}=\frac{\left(  b-c\right)
^{3}\chi(a)+\left(  c-a\right)  ^{3}\chi(b)+\left(  a-b\right)  ^{3}\chi
(c)}{\left(  b-a\right)  \left(  c-b\right)  \left(  c-a\right)  }
\label{energabc}%
\end{equation}
where
\[
\chi(x)\equiv\sum_{m=1}^{\infty}h_{m}x^{m}.
\]

Let us now introduce the hedgehog ansatz
\begin{equation}
U_{0}(\mathbf{{x})=}\exp\left[  i\tau_{i}\hat{x}_{i}F(r)\right]  ,
\label{eq:hegehog}%
\end{equation}
where the $\tau_{i}$ are the three Pauli matrices and the $\hat{x}_{i}$ are
the three components of a radial unit vector. This form (\ref{eq:hegehog}) is
known to minimize the static energy of $B=1$ skyrmions. The exact solution
requires the numerical computation of the chiral angle $F(r)$. Using this
ansatz for general Lagrangians (\ref{eq:L}), we may write the static energy
density $\mathcal{E}$ which is at most quadratic in $F^{\prime}$
\begin{equation}
\mathcal{E}=3\chi(a)+(b-a)\chi^{\prime}(a), \label{eq:Densite_E}%
\end{equation}
since the hedgehog ansatz, $a=c\equiv\sin^{2}F/r^{2}$ and $b\equiv(F^{\prime
})^{2}$ . The Skyrme model corresponds to the particular case $\chi
(x)=\chi_{S}(x)\equiv x+x^{2}/2$. Minimizing (\ref{eq:Densite_E}) using the
Euler-Lagrange equation leads to%
\begin{equation}
0=\chi^{\prime}(a)\left[  F^{\prime\prime}+2\frac{F^{\prime}}{r}-2\frac{\sin
F\cos F}{r^{2}}\right]  +a\chi^{\prime\prime}(a)\left[  -2\frac{F^{\prime}}%
{r}+(F^{\prime})^{2}\frac{\cos F}{\sin F}+\frac{\sin F\cos F}{r^{2}}\right]  .
\label{eq:chiral_equation}%
\end{equation}
This differential equation is at most of degree two, thus computationaly
tractable. Such a requirement was first proposed in \cite{Marleau1990}. In
fact, the Skyrme model or any model made up of a linear combination of
$\mathcal{L}_{1},\mathcal{L}_{2}$ and $\mathcal{L}_{3}$ satisfy this condition
so in a sense, the models we are interested in are their natural extensions.
As we shall see, the SA algorithm minimizes energy functionals directly, and
therefore there is no need to invoque Euler-Lagrange formalism and to write
down the corresponding differential equations.

The hedgehog ansatz is particularly useful for the $B=1$ solutions, which have
been shown to possess spherical symmetry. This is not the case for higher
topological sectors whose symmetries are fortunately well approximated by the
rational maps ansatz.

\section{\label{sec:RatMaps}Rational maps ansatz for skyrmions}

It has been shown \cite{Battye19971, Battye2001} by numerical work that the
static Skyrme solitons of charge $1<B\leq22$ are not radially symmetric. In
that context, the hedgehog ansatz needs to be replaced by a more general
ansatz in order to reproduce the specific angular distributions of
multi-skyrmions. Hopefully, there exists an ansatz based on rational maps
\cite{Houghton1998} which constitutes a very good approximation for $B>1$ skyrmions.

One of the aims of this paper is to evaluate the static energy of $B=1$ to
$B=4$ skyrmions using the rational maps ansatz, and then compare these results
with the exact numerical ones we obtain for several chiral models. This
strategy will give us keen information on the reliability of the ansatz for
certain extended Skyrme models, which is currently unknown, in particular for
a model comprising an order eight extension. A correspondence between exact
and rational maps solutions would then allow one to determine the symmetries
of an exact numerical solution from the analysis of the matching rational map.

A priori there seems to be no reason why the rational map ansatz should
provide good approximation for solutions of extended models unless these
models are numerically very close to the Skyrme model. However, looking at the
energy density (\ref{energabc}), one realizes that when the two invariants $a$
and $c$ (which are identified with the angular distribution in our case) are
equal, $\mathcal{E}$ becomes linear in $b.$ Otherwise the contribution of each
term $\mathcal{E}_{m}$ would have been of order $b^{m-2}.$ Since rational maps
are conformal maps they preserved the relation $a=c$ in which case only terms
in the energy density which are at most linear\ in $b$ survive and presumably
this would correspond to an energically favoured configuration. So one might
conjecture that the soliton solutions for the class of models defined in
\cite{Marleau1990} are well represented by the rational map ansatz or that
rational map solutions would remain well suited approximation for extended
models as long as they belong to this particular class of models. This is the
main motivation to compare rational maps inspired solutions with the exact
numerical solutions for such models.

To briefly review the rational maps ansatz, let us introduce the coordinates
$(r,z,\bar{z})$ parametrizing a point $\mathbf{x}$ in $R^{3}$, where
$r=|\mathbf{{x}|}$ is the distance from the origin, and $z=\tan(\theta
/2)\exp(i\phi)$ and $\bar{z}$, its complex conjugate, encode the angular
dependance. Following Houghton, Manton and Sutcliffe \cite{Houghton1998}, we
approximate the structure of skyrmions by a real chiral angle $F(r)$,
satisfying the boundary conditions $F(0)=\pi$ and $F(\infty)=0$, and an ansatz
$U(r,z)$ based on rational maps
\begin{equation}
R(z)=\frac{p(z)}{q(z)},
\end{equation}
written in terms of polynomials $p$ and $q$ having no common factors. The
degree $N=\mathrm{max}[\mathrm{deg}(p),\mathrm{deg}(q)]$ of the map
corresponds to the baryonic number $B$ of the soliton. Given such maps, the
Skyrme fields take the form
\begin{equation}
U(r,z)=\exp\left[  \frac{iF(r)}{1+|R|^{2}}~%
\begin{pmatrix}
1-|R|^{2} & ~2\bar{R}\\
2R & ~|R|^{2}-1
\end{pmatrix}
\right]  . \label{eq:ansatz_maps}%
\end{equation}
Substituting (\ref{eq:ansatz_maps}) in the energy density of the Skyrme model
and integrating over the angular degrees of freedom, we find the energy
functional
\begin{equation}
E_{S}=-\int\mathcal{L}_{S}~d^{3}x=4\pi\int\left(  r^{2}F^{\prime}%
(r)^{2}+2N(F^{\prime}(r)^{2}+1)\sin^{2}F(r)+\mathcal{I}~\frac{\sin^{4}%
F(r)}{r^{2}}\right)  dr, \label{eq:ES_maps}%
\end{equation}
where $N=B$ is the topological number
\begin{equation}
N\equiv\frac{1}{4\pi}\int\left(  \frac{1+|z|^{2}}{1+|R|^{2}}\left\vert
\frac{dR}{dz}\right\vert \right)  ^{2}\frac{2i~dzd\bar{z}}{(1+|z|^{2})^{2}}%
\end{equation}
and $\mathcal{I}$ denotes the integral
\begin{equation}
\mathcal{I}\equiv\frac{1}{4\pi}\int\left(  \frac{1+|z|^{2}}{1+|R|^{2}%
}\left\vert \frac{dR}{dz}\right\vert \right)  ^{4}\frac{2i~dzd\bar{z}%
}{(1+|z|^{2})^{2}}~. \label{eq:int_I}%
\end{equation}
One recognizes the angular distribution of the baryonic density
\[
\rho(z,\bar{z})=\left(  \frac{1+|z|^{2}}{1+|R|^{2}}\left\vert \frac{dR}%
{dz}\right\vert \right)  ^{2}%
\]
and the element of solid angle $\sin\theta d\theta d\phi=\frac{2i~dzd\bar{z}%
}{(1+|z|^{2})^{2}}$. The minimization of the energy (\ref{eq:ES_maps})
requires that we first find the rational map that minimizes (\ref{eq:int_I})
for a given degree $N$ and then find the chiral angle $F(r)$ by minimizing the
energy in (\ref{eq:ES_maps}). For the Skyrme model, the rational maps
approximation gives an energy accurate within $1\%$ or $2\%$ compared with
exact numerical results, except for the $B=1$ radially symmetric skyrmion
where the ansatz yields the exact result.

To study extensions of the Skyrme model in topological sectors $B>1$ with the
rational maps ansatz, we need to modify the expression for the energy density
(\ref{eq:Densite_E})
\[
\mathcal{E}=\sum_{m=1}^{\infty}h_{m}a^{m-1}\left[  3a+m(b-a)\right]  ,
\]
with $a,b$ and $c$ now given by the more general expressions
\begin{align}
a  &  =c=a_{F}\rho(z,\bar{z})\qquad\text{with}\qquad a_{F}=\frac{\sin^{2}%
F}{r^{2}}\\
b  &  =(F^{\prime})^{2}%
\end{align}
where the angular dependence is no longer trivial. Note that the angular
distribution of the energy density is entirely included in $\rho(z,\bar{z})$,
and powers of $a$ can take into account the angular distribution of any given
rational maps ansatz. Working with the notation $\chi(x)=\sum_{m=1}^{\infty
}h_{m}x^{m}$, the static energy of the skyrmion is then written
\begin{equation}
E=4\pi\left(  \frac{F_{\pi}}{2\sqrt{2}e}\right)  \int_{0}^{\infty}r^{2}%
dr\int\frac{2i~dzd\bar{z}}{4\pi(1+|z|^{2})^{2}}~[3\chi(a_{F}\rho(z,\bar
{z}))+(b-a_{F}\rho(z,\bar{z}))\chi^{\prime}(a_{F}\rho(z,\bar{z}))].
\label{eq:Mass_chi}%
\end{equation}
The integrals over the angular degrees of freedom present in
(\ref{eq:Mass_chi}) are not trivial in general. The angular dependence is
however easy to isolate when $\chi(x)$ is a polynomial, in which case we need
to evaluate
\begin{equation}
I_{m}^{N}\equiv\frac{1}{4\pi}\int\frac{2i~dzd\bar{z}}{(1+|z|^{2})^{2}}\left(
\rho(z,\bar{z})\right)  ^{m}. \label{eq:INM}%
\end{equation}
All the angular dependence for a given choice of rational map is then
contained in the integrals (\ref{eq:INM}), leading to the following expression
for the static energy (or the mass) of the skyrmion
\begin{equation}
E=4\pi\left(  \frac{F_{\pi}}{2\sqrt{2}e}\right)  \int_{0}^{\infty}%
r^{2}dr~(3\chi_{1}(a_{F})+b\chi_{2}^{\prime}(a_{F})-a\chi_{1}^{\prime}(a_{F}))
\label{eq:Mass}%
\end{equation}
where
\begin{align}
\chi_{1}(a)  &  =\sum_{m=1}^{\infty}h_{m}a_{F}^{m}I_{m}^{N}\\
\chi_{2}(a)  &  =\sum_{m=1}^{\infty}h_{m}a_{F}^{m}I_{m-1}^{N}.
\end{align}
Since we want to compare general solutions with those of the Skyrme model, we
use as a starting point the rational maps that minimize the energy for the
Skyrme model. Thus, for the topological sectors $N=2$, $N=3$ and $N=4$, the
symmetries of the rational maps ansatz are, respectively, toroidal,
tetrahedral and cubic, corresponding to the maps
\begin{equation}
R(z)=z^{2},
\end{equation}%
\begin{equation}
R(z)=\frac{\sqrt{3}az^{2}-1}{z(z^{2}-\sqrt{3}a)},
\end{equation}
where $a=\pm i$,
\begin{equation}
R(z)=\frac{z^{4}+2\sqrt{3}iz^{2}+1}{z^{4}-2\sqrt{3}iz^{2}+1}.
\end{equation}
According to (\ref{eq:Mass}), we need first here to evaluate the angular
integrals (\ref{eq:INM}) for $1\leq m\leq4$ and $1\leq N\leq4$ in order to
proceed with the minimization procedure involved in the determination of
$F(r)$. In the case of spherical symmetry $R(z)=z$, it is trivial to show that
$I_{m}^{1}=1$ for all $m$. The numerical values of the integrals $I_{m}^{N}$
for the topological sectors $N=2$, $N=3$ and $N=4$ are presented in table
\ref{tab:Int_angulaires_num}.

\begin{table}[ptb]
\begin{center}%
\begin{tabular}
[c]{|c|c|c|c|}\hline
$I_{m}^{N}$ & $N=2$ & $N=3$ & $N=4$\\\hline
$m=1$ & $2$ & $3$ & $4$\\
$m=2$ & $5.808$ & $13.577$ & $20.650$\\
$m=3$ & $19.025$ & $75.997$ & $118.244$\\
$m=4$ & $65.901$ & $484.868$ & $719.720$\\\hline
\end{tabular}
\hspace{1cm}
\end{center}
\caption{Numerical values of the angular integrals $I_{m}^{N}$ defined in
(\ref{eq:INM}) for rational maps of degree $N=2$ (toroidal symmetry), $N=3$
(tetrahedral symmetry) and $N=4$ (cubic symmetry).}%
\label{tab:Int_angulaires_num}%
\end{table}

\section{\label{sec:Positivity}Positivity constraints}

As a guiding principle in our choice of models (or weights $h_{m}$), we
require that the energy density of the models constructed out of linear
combinations of $\mathcal{L}_{m}$ be positive. Since one of the purposes of
this work is to compare exact solutions with the rational maps approximation
in order to easily identify the symmetries of the solutions, a second
constraint must be imposed on the models. As illustrated in the last section,
the rational maps approximation is calculable only for models with a finite
number of $\mathcal{L}_{m}$ where the angular dependence can be integrated
out. For calculational purposes, we shall therefore limit the analysis here to
the most simple extended models: order-six and order-eight Lagragians, i.e.
$m<5$ in (\ref{eq:L}), although we did briefly explore other higher-order
models. Finally, the solutions should be stable against a scale
transformation, i.e. there must be a mechanism which prevents the skyrmions
from shrinking to zero size. These three simple requirements prove themselves
to be quite restrictive in practice, narrowing substantially the possibilities
for the higher weights, in particular $h_{4}$ in the order-eight Lagrangian.
In fact, contrary to what one might suspect, taking positive $h_{m}$ does not
guarantee the existence of a positive energy solution. Proceeding first by
trial and error to set the value of $h_{4}$, we encountered systematically non
positive energy densities in most of our numerical calculations. Indeed it was
quite difficult to obtain models with a positive energy density when the
$\chi(a)$ series has a polynomial form and its last term is of order eight or
more in derivatives of the pion fields.

So, we have to take a closer look at the positivity constraint. Jackson, Weiss
and Wirzba \cite{Jackson1991} first established a set of rules for extended
models. For positivity, they must obey
\[
\chi^{\prime}(x)\geq0~\text{ and }~3\chi(x)-x\chi^{\prime}(x)\geq0~\text{ for
}~x\geq0 ,
\]
which are sufficient but not necessary conditions for positivity and may be
accompagnied by other conditions such as chiral symmetry restauration
\cite{Jackson1991} to construct viable models. These constraints however only
apply in the context of the $B=1$ spherically symmetric hedgehog solution.
This is clearly not appropriate for our numerical analysis where no specific
ansatz or symmetries are imposed.

Let us recall the most general form of the energy density for a model defined
by a function $\chi$
\begin{equation}
\mathcal{E}=\frac{\left(  b-c\right)  ^{3}\chi(a)+\left(  c-a\right)  ^{3}%
\chi(b)+\left(  a-b\right)  ^{3}\chi(c)}{\left(  b-a\right)  \left(
c-b\right)  \left(  c-a\right)  } \label{Eabc}%
\end{equation}
Note that both the hedgehog and rational maps ansatz assume uniform angular
energy distribution which means that in these cases, two of the invariants are
equal. Without loss of generality, we take
\[
0\leq a\leq b\leq c.
\]
$\mathcal{E}$ in (\ref{Eabc}) is positive if%
\begin{equation}
\chi(b)\geq\alpha^{3}\chi(a)+\left(  1-\alpha\right)  ^{3}\chi(c) \label{pos1}%
\end{equation}
where $\alpha=\frac{c-b}{c-a}$ and $0\leq\alpha\leq1$. But if (\ref{pos1}) is
true, then%
\[
\chi(b)\geq\left(  \alpha\chi^{\frac{1}{3}}(a)+\left(  1-\alpha\right)
\chi^{\frac{1}{3}}(c)\right)  ^{3}\geq\alpha^{3}\chi(a)+\left(  1-\alpha
\right)  ^{3}\chi(c)
\]
or%
\[
\chi^{\frac{1}{3}}(b)\geq\alpha\chi^{\frac{1}{3}}(a)+\left(  1-\alpha\right)
\chi^{\frac{1}{3}}(c)~.
\]
The last condition is obeyed if $\chi^{\frac{1}{3}}(x)$ is a positive and
concave function over an interval which includes $a,b$ and $c$. There are
other instances where the condition is satified, e.g. if $\chi^{\frac{1}{3}%
}(x)$ is a negative and convex function over the interval including $a,b$ and
$c$. Note that $\chi(a),\chi(b)$ and $\chi(c)$ may not have the same sign in
general, and for arbitrary $a,b,c$ checking positivity analytically is not
possible. Here, since the first term in $\chi(x)$ corresponds to the nonlinear
$\sigma$ term, $h_{1}$ is taken to be positive and at least for small values
of $x,$ $\chi(x)$ is positive and $\chi^{\frac{1}{3}}(x)$ must be concave. So
we shall explore models where $\chi^{\frac{1}{3}}(x)$ remains positive and concave.

It is trivial to show that all order-six models with positive weights
$h_{1},h_{2}$ and $h_{3}$ obey this positivity constraint. Our numerical
analysis was indeed free of problems related to energy positivity. As
mentioned before, the problem of energy positivity began to arise when we
added a term of order eight. It was nonetheless possible to construct a model
that satisfies the constraint. An appropriate value for $h_{4}$ was first set
by trial and error, checking numerically that the energy density was positive
everywhere. The models we study in this work are defined by the functions
$\chi(x)\equiv\sum_{m=1}^{\infty}h_{m}x^{m}$ with
\begin{equation}%
\begin{tabular}
[c]{rl}%
$\text{Skyrme}:$ & $\chi_{S}(x)=x+\frac{x^{2}}{2}$\\[0.1cm]%
$\text{Order six}:$ & $\chi_{O6}(x)=x+\frac{x^{2}}{2}+\frac{x^{3}}{6}%
$\\[0.1cm]%
$\text{Order eight}:$ & $\chi_{O8}(x)=x+\frac{x^{2}}{2}+\frac{x^{2}}{6}%
-\frac{x^{4}}{240}$%
\end{tabular}
\ \label{order246}%
\end{equation}
or the Lagrangians $\mathcal{L}\equiv\sum_{m=1}^{\infty}h_{m}\mathcal{L}_{m}$
where $\mathcal{L}_{4}=-\frac{4}{3}\mathcal{L}_{1}\mathcal{L}_{3}%
+\mathcal{L}_{2}^{2}$.

Using the concavity constraint on $\chi^{\frac{1}{3}}(x)$, it is now easy to
see that $h_{4}$ must be negative in an order-eight model. However in
principle, the highest order term is responsible for the stability against a
scale transformation, and a negative $h_{4}$ in this model would allow the
soliton to shrink to zero size unless another mechanism brings stability. It
turns out that our numerical approach provides such a mechanism naturally. The
solutions are found by discretizing space and by construction the soliton size
cannot be smaller than the lattice spacing while having a finite baryon
number. So what we are looking for is a finite size solution that minimizes
energy. The solutions may not be a global minimum in energy for the
order-eight model but they could be very close to solutions of more physically
relevant models. For example, the eight-order model may be seen as an
approximation of the model having the rational form defined by%
\begin{equation}
\chi_{R}(x)=x+\frac{x^{2}}{6}\frac{120+43x}{40+x} \label{allorder}%
\end{equation}
or the Lagrangian%
\[
\mathcal{L}_{R}(x)=\mathcal{L}_{1}+\frac{x^{2}}{6}\frac{120+43x}{40+x}%
\]
which in turns has positive energy solutions that are stable against scale
transformations. Unfortunately, the solutions for such a model cannot be
compared directly with rational maps solutions and this is why we only
considered finite polynomial form for $\chi$ in the first place. For
comparison, the solutions for the model $\chi_{R}$ will also be presented.

\section{\label{sec:SimAnn}Simulated annealing algorithm}

Despite the numerous attractive features that characterize topological nuclear
theories like the Skyrme model, it is not an easy task to extract their
solutions. This drawback is partly due to the nonlinear structure of such
models. Therefore, one must rely on numerical methods to accomplish the
minimization of energy functionals in these cases.

In the following section, we first describe how SA works by presenting its
structure and its most important features in a problem independent way. Then,
we review our implementation of the SA algorithm, inspired by the guidelines
stated in \cite{Hale2000}, for two different strategies we have used to solve
the Skyrme model and some of its generalizations. In the one dimensional case,
the angular dependence of the energy functional is integrated out with an
appropriate rational maps ansatz, then the SA algorithm is used to find the
chiral angle function $F(r)$ leading to the minimal value for the static
energy of the skyrmion. This is by no mean the most straightforward numerical
technique to get the exact solution nor is it the most commonly used or the
fastest but it will provide here an estimate of the kind of accuracy the SA
algorithm can achieve. Then, the exact solutions will be evaluated using a
three dimensional implementation of the SA algorithm.

\subsection{\label{sec:Metropolis}Simulated annealing and the Metropolis
algorithm}

The SA algorithm is an optimization tool that proves to be usefull in several
problems. In the case of a classical field theory, we are interested in the
field configuration $\phi_{min}(x)$ which gives the lowest value of an energy
functional
\begin{equation}
E=\int_{M}d^{n}x~\mathcal{E}(x,\phi(x),\phi^{\prime}(x)),
\label{eq:general_functional}%
\end{equation}
where $\mathcal{E}$ is the energy density. The solution $\phi_{min}(x)$ is
found over all the configurations $\phi(x)$ satisfying some topological
boundary conditions on the manifold $M$. Using the Euler-Lagrange formalism,
the minimal value of $E$ would correspond to the solution of a differential
equation generated from (\ref{eq:general_functional}). Conversely, working
with the SA algorithm does not require to write and solve differential
equations which makes it particularly useful if these turn out too complex.
The field configuration $\phi_{min}(x)$ is the result of a Monte-Carlo
process, an iterative improvement technique.

In its simplest form, the iterative process starts with a given system in a
known configuration with respect to a cost function like
(\ref{eq:general_functional}). Each step of the process corresponds to a
random rearrangement operation applied to some parts of the system. If the
rearrangement of the configuration leads to a lower value of the cost
function, it is accepted and the next iterative step starts with this new
configuration. On the other hand, if the rearranged configuration makes the
value of the cost function rise, it is rejected and we put the system back in
its former configuration. Iterations are made until no further downhill
improvements on the cost function can be found. This technique is not
completely reliable though, because it is frequent to observe searches get
stuck in a local minimum of the cost function, rather than the global one we seek.

In 1983, Kirkpatrick, Gellat and Vecchi \cite{Kirkpatrick1983} introduced the
SA algorithm in the framework of iterative processes. Their aim was to improve
the convergence towards the global minimum of a cost function. Using the
analogy between the cost function and a solid that is slowly cooled down, they
proposed to add a temperature to the problem. The idea is simple. First, heat
the system to a certain temperature $T$ and run the Metropolis algorithm to
bring the system into thermal equilibrium. Then, allow the temperature to
decrease slowly insuring thermal equilibrium at each step during the cooling
process. In the limit of a sufficiently low temperature, the solid will reach
its ground state. Equivalently, one will find the minimal energy configuration
$\phi_{min}(x)$ of the functional (\ref{eq:general_functional}) in this limit.

\begin{figure}[ptb]
\begin{center}
\includegraphics[scale=0.5]{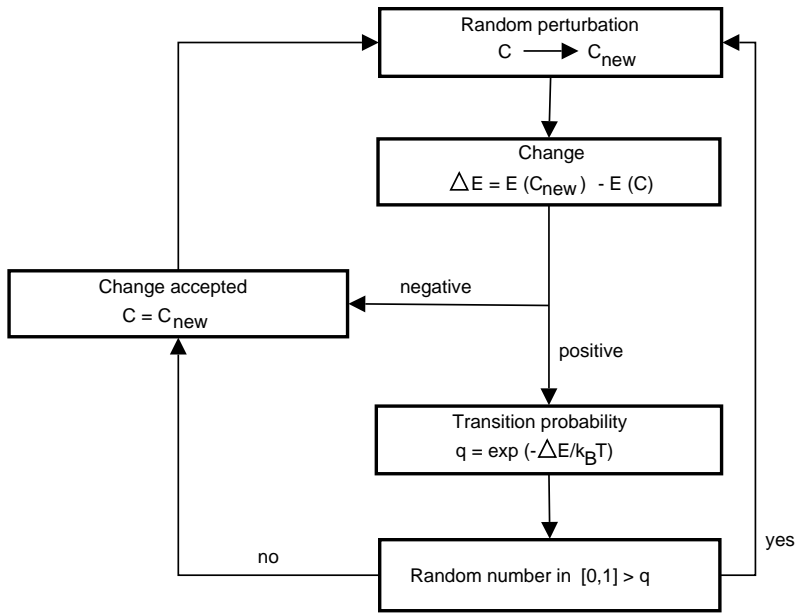}
\end{center}
\caption{Logical scheme of the Metropolis algorithm.}%
\label{fig:Schema-Logique}%
\end{figure}

The Metropolis algorithm \cite{Metropolis1953} was originally developped in
1952 to accurately simulate a group of atoms in equilibrium at a given
temperature. The logical scheme of this algorithm is shown in figure
\ref{fig:Schema-Logique}. At each step of the iterative process, a random
rearrangement of the configuration is made leading to a modification $\Delta
E$ of the cost function. If the cost is lowered by this rearrangement $\Delta
E\leq0$, it is accepted and the next iterative step starts with the new
configuration. If the rearrangement increases the cost $\Delta E>0$, it is not
automatically rejected. Instead, a probability factor $q=\exp(-\Delta
E/k_{B}T)$ constructed out of the cost modification $\Delta E$ is compared
with a random number in the interval $[0,1]$. If the factor $q$ is greater
than the random number, the new configuration is accepted, otherwise it is
rejected. So, it becomes statistically more difficult for a rearranged
configuration with $\Delta E>0$ to be accepted as the temperature $T$
decreases. This scheme can be applied directly to a cost function like
(\ref{eq:general_functional}).

Let us now state a number of important aspects to keep in mind when
implementing an SA algorithm. (i) The initial temperature must be chosen with
care. If one starts a simulation with the temperature set too high, the
soliton may unwind. Also, the overall running time will be longer because of
the increased number of cooling steps then needed. On the other hand, if the
initial temperature is set too low, the system may stabilize in a local
minimum when it is sufficiently cooled down. (ii) The cooling schedule must be
smooth enough. It has been shown by Geman and Geman \cite{Geman} that if the
temperature decreases according to a logarithmic rule, the convergence to the
global minimium of the system is guaranteed. However, faster cooling schemes
can produce reliable results too. (iii) It is important that the initial guess
for the field configuration possesses the winding number of the solution we
seek. If it is not the case, the formation of isolated lumps of baryonic
density is likely to occur as the simulation progresses. In our three
dimensional simulations, we initialize the lattice with an appropiate degree
rational map. (iv) The conservation of the baryonic number must be imposed
explicitly in three dimensional simulations. This is done by adding a Lagrange
multiplier to the action. This object tends to reject a trial configuration
that compromises the quality of the topological integer, even if this
configuration implies a lower value of the cost function. In the one
dimensional case, this is not an issue since the conservation of the baryonic
number is assured by the boundary conditions on the chiral angle $F(r)$.

The details concerning the sampling method we use and our criterion for the
determination of thermal equilibrium will be discussed shortly.

\subsection{\label{sec:SA1D}One dimensional simulated annealing and rational
maps}

In our one dimensional approach, we first integrate out the angular dependence
of the skyrmion using the rational maps ansatz. Then, SA finds the chiral
angle $F(r)$ leading to a minimal value of the energy functional for
generalized Skyrme models (\ref{eq:Mass}). To do this, we need a lattice that
represents the radial degree of freedom $r$ in a discretized manner. In this
context, the conservation of the topological integer $N$ is guaranted by the
boundary conditions $F(0)=\pi$ and $F(\infty)=0$ and the degree $N$ of the
rational map $R(z).$

To minimize the energy functional of a model, we need to compute $F(r)$ on the
lattice. The energy and the baryonic number are evaluated halfway between two
adjacent points of the lattice, $r_{i}$ and $r_{i+1}$. The value that the
chiral angle and its derivative take there are given by
\begin{align}
F(r_{i+1/2})  &  =\frac{F(r_{i})+F(r_{i+1})}{2}~,\\
\frac{dF(r_{i+1/2})}{dr}  &  =\frac{F(r_{i+1})-F(r_{i})}{dr}~,
\end{align}
where $dr=r_{i+1}-r_{i}$. A \emph{cell} represents the space between two
adjacent lattice points. One assigns to each cell a value of energy and
baryonic density as calculated at its center $r_{i+1/2}$.

SA relies on an iterative random process. To ensure that the sampling in space
of field configurations is performed correctly, we must efficiently rearrange
$F(r)$ during the simulation. We choose to perturbate randomly just one
discrete value of $F(r)$ at each iteration, namely the one corresponding to
the point $r_{i}$. Doing so, the energy and baryonic density of the two cells
sharing the point $r_{i}$ are modified. This alteration of $F(r_{i})$ is then
accepted or rejected according to the Metropolis algorithm. Explicitely, a
random change of $F(r_{i})$ takes the form
\begin{equation}
F(r_{i})\longrightarrow F(r_{i})+A\cos(2\pi n),
\end{equation}
where $A$ is the maximal amplitude of the change and $n$ is a random number in
the interval $[0,1]$. An appropriate and efficient sampling will occur if $A$
is adjusted dynamically during the simulation so that $50\%$ of the new
configurations are rejected. This strategy implies that $A$ decreases linearly
with temperature $T$.

To implement an SA algorithm properly, thermal equilibrium must be reached at
each temperature of the cooling scheme. The strategy we use is based on a
statistical chain composed of a given number of iterations. The chain has a
length set to 10 times the number of points on the lattice. This has the
following statistical consequence. When the algorithm has done a number of
random perturbations corresponding to the length of the chain, the value of
$F(r)$ has been modified approximately 10 times at each point of the lattice.
To state that thermal equilibrium is reached, we compare the lowest energy
value encountered in the chain $E_{\mathrm{min}}$ and the sum of the energies
obtained at each iteration of the chain $\sum_{\mathrm{tries}}E_{\mathrm{try}%
}$ to their equivalent in the previous chain. If these two values are higher
in the present chain than in the previous, thermal equilibrium has been
reached and the temperature may be lowered. In other words, another chain must
be started at the same temperature as long as

\begin{itemize}
\item[$\bullet$] $E_{min}$ of the \emph{present} chain $<$ $E_{min}$ of the
\emph{previous} chain or,

\item[$\bullet$] $\sum_{\mathrm{tries}}E_{\mathrm{try}}$ of the \emph{present}
chain $<$ $\sum_{\mathrm{tries}}E_{\mathrm{try}}$ of the \emph{previous} chain.
\end{itemize}

\begin{figure}[ptb]
\begin{center}
\includegraphics[scale=0.4, angle=270]{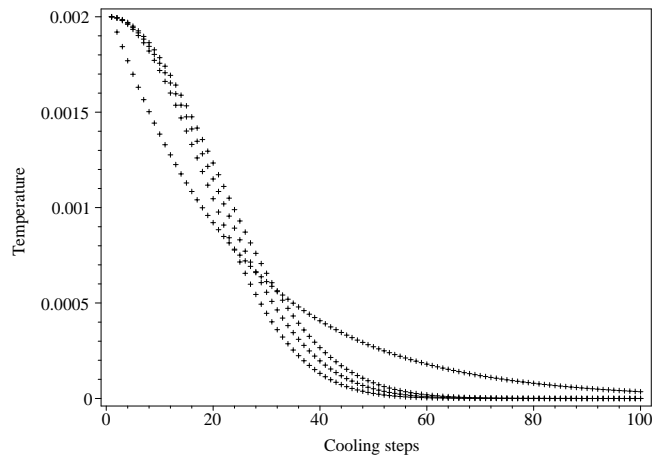}
\end{center}
\caption{Temperature as a function of the cooling steps according to four
different schemes. The curve which is distinct from the three others
corresponds to a logarithmic scheme, i.e. of the form $T_{i}=0.95~T_{i-1}$.
The other curves represent cooling schemes possessing the structure
$T_{i}=T_{i-1}(X-i)/X$, where $X$ is fixed to $300$, $350$ and $400$,
respectively from bottom to top.}%
\label{fig:Coolings}%
\end{figure}

Every time thermal equilibrium is achieved, the temperature is decreased
according to a cooling schedule. In our one dimensional simulations, we use a
logarithmic cooling because the computational cost is not an issue. This kind
of schedule decreases the temperature by a fixed ratio at each step, for
example $T_{i}=0.95~T_{i-1}$, presented on figure \ref{fig:Coolings}. In
practice, faster cooling schemes can be used without negative impact on the
quality of the global minimum obtained. Experience has shown that such a
faster scheme can take the form $T_{i}=T_{i-1}(X-i)/X$, where $X$ is a
constant factor, which still leads to reliable results. We have set the
initial temperature to $T_{\mathrm{ini}}=1$ in all our one dimensional
simulations. This value was obtained by trial and error while comparing the
quality of the minimization.

To speed up the minimization process, we have even implemented an adaptive
lattice. The simulation is started with a reasonably low number of points and
a chiral angle which is a linear interpolation between $F(0)=\pi$ and
$F(150)=0$. Typically, 100 points are used to cover the lattice that has 150
units of length $2\sqrt{2}/(eF_{\pi})$. When a global minimum is found,
additional lattice points are added where the energy is concentrated and the
temperature is set to its initial value $T_{\mathrm{ini}}=1$ before SA starts
again. This procedure is repeated until the desired precision on the solution
for $F(r)$ is achieved, corresponding to a lattice of approximately 650
points. However, this adaptive procedure is only practical in the one
dimensional case, where the computational cost is relatively low. In three
dimensional SA, computational time and computer memory considerations become
important constraints. So, cooling the system several times to achieve the
most convenient lattice spacing would not constitute an advantage in that case.

\subsection{\label{sec:SA3D}Three dimensional simulated annealing}

The three dimensional SA algorithm is used to find exact solitonic solutions.
This approach is a full field simulation where no constraints on the
symmetries of the solution are imposed. Here are some important aspects of our
three dimensional implementation.

The lattice we use is composed of $80\times80\times80$ points equally spaced
by $0.12$ units of length $2\sqrt{2}/(eF_{\pi})$. This lattice spacing is
somewhat large, but is a good compromise in regard to reducing the
computational time. The cartesian axes go from $-4.8$ units to $4.8$ units in
the $x$, $y$ and $z$ coordinates. This finite volume can cause undesirable
numerical effects on the edge of the lattice. To significantly reduce these,
we use a lattice with periodic boundaries. In this case the skyrmion interacts
with itself over the boundaries, but its structure is not altered. The
skyrmion only suffers a slight increase in energy.

\begin{figure}[ptb]
\begin{center}
\includegraphics[scale=0.4, angle=270]{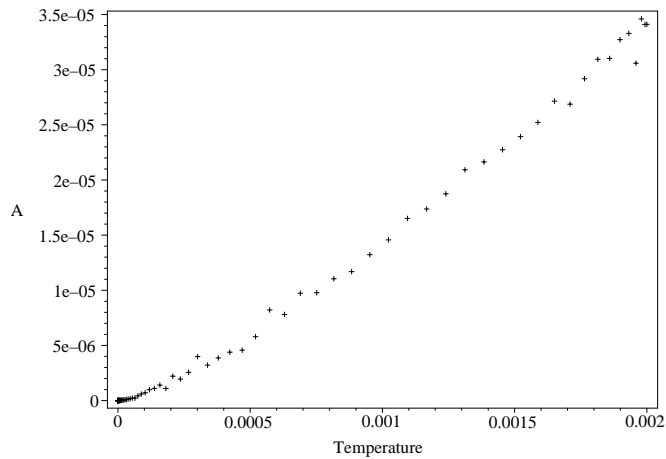}
\end{center}
\caption{Maximal amplitude $A$ of the random perturbations done on 3D field
vectors as a function of temperature for the Skyrme model.}%
\label{fig:A_fct_T_3D}%
\end{figure}

Because of topology, each lattice point is characterized by a four-component
field $\phi_{m}$ which satisfies $\phi_{m}\phi_{m}=1$. As a consequence,
perturbing the configuration by random rotations must also preserve the length
of $\phi_{m}$, as explained in \cite{Hale2000}. These rotations are
parametrized in terms of three random Euler angles. For computational
purposes, we only consider small perturbations for one of the angles by
limiting the random values to an upper bound $A$. Covering the whole solid
angle is ensured by multiple rotations if needed. We adjust dynamically the
parameter $A$ during the cooling process so that $60\%$ of the perturbations
performed on the field $\phi_{m}$ are rejected. Starting with $T_{\mathrm{ini}%
}=0.02$ and using a cooling schedule of the form $T_{i}=T_{i-1}(300-i)/300$,
where $i$ is a cooling step, the amplitude $A$ drops linearly with temperature
(figure \ref{fig:A_fct_T_3D}). This ensures that the rearrangements will not
be too large as the temperature decreases, thus respecting the SA scheme.

\begin{figure}[ptb]
\begin{center}
\includegraphics[scale=0.16, angle=270]{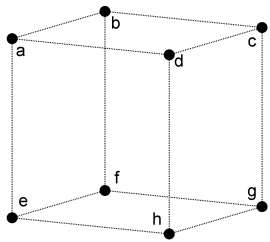}
\end{center}
\caption{A cubic cell from the 3D lattice. This cell has a volume of
$(0.12)^{3}$ units $(2\sqrt{2}/(eF_{\pi}))^{3}$. We identify its vertices with
indices for further reference. The $x$, $y$ and $z$ axes are assigned in the
directions of the segments $\overline{eh}$, $\overline{ef}$ and $\overline
{ea}$ respectively.}%
\label{fig:cell_3D}%
\end{figure}

To sample the fields on the lattice, we use again the concept of cell (figure
\ref{fig:cell_3D}). An iteration in the statistical chain starts with the
random selection of a cell. Then, each of the eight vertices (lattice points)
of the cell undergoes a rotation of the field. These rearrangements modify the
energy and baryonic densities calculated at the center of the current cell and
at the center of the 26 neibourghing cells, each of them having at least one
lattice point in common with the randomly modified cell. We then use the
Metropolis criterion to analyse the effect of the rotations on the energy of
the 27 cells subsystem. If the acceptance criterion is confirmed, the changes
affecting the eight lattice points are all accepted otherwise they are all rejected.

To compute the energy and baryonic densities at the center of a cell, one must
rely on a linear interpolation using the value of the field at each of the
eight vertices of the cell. Then, the energy density is evaluated at the
center of the cell and the contribution to the total energy is obtained by
mutiplying the density by the volume of the cell, $(0.12)^{3}$. Unfortunately,
the linear interpolation does not preserve the relation $\phi_{m}\phi_{m}=1$,
so we have to scale the fields following the interpolations. For example, the
derivative at the center of a cell in the $x$ direction, refering to figure
\ref{fig:cell_3D}, is given by
\begin{equation}
\left.  \frac{\partial\phi}{\partial x}\right\vert _{\text{~center}}=\frac
{1}{0.12}~\left\{  \text{Scaled}\left(  \frac{1}{4}~[\phi(c)+\phi
(d)+\phi(g)+\phi(h)]\right)  -\text{Scaled}\left(  \frac{1}{4}~[\phi
(a)+\phi(b)+\phi(e)+\phi(f)]\right)  \right\}  ~.\nonumber
\end{equation}

In our 3D simulations, we need to introduce a procedure to ensure that the
baryonic number of the field configuration is conserved throughout the cooling
process. A convenient way is to add a Lagrange multiplier to the action. This
piece filters only appropriate random perturbations generated by the
Metropolis algorithm. The Lagrange multiplier will tend to reject
rearrangements compromising the quality of the calculated topological integer,
regardless of their effect on the value of the energy functional being
minimized. To help the convergence of SA towards the global minimium of the
desired topological sector, we initialize the lattice with a field
configuration that is already in the sector of interest. We use a rational map
ansatz of adequate degree to generate an initial field configuration. In fact,
the precise form of the initialization ansatz does not matter, as long as its
degree $N=B$. For simplicity, we use the ansatz $R(z)=z^{N}$. Such a
configuration possesses radial symmetry for $N=1$ and toroidal symmetry for
$N>1$.

\section{\label{sec:Results}Numerical results and discussion}

First, let us take a look at our results for the static energies and baryonic
numbers presented in table \ref{Sk_vs_O6_O8}. For the models considered in
this work, we note that adding a positive (negative) weight $h_{m}$ leads to
an increase (decrease) in the static energy of the skyrmion which depends on
the size of $h_{m}$. For example, considering that $h_{4}$ is small compared
to $h_{3}$, such a feature could explain the slight difference between the
energies of the order-six and the order-eight models. Also, comparing the
ratios $E_{N>1}/E_{1}$, multi-skyrmions solutions of extended models appear to
be more stable than for the Skyrme model, for a greater part of their energy
is used to create solitonic bound states. However, we cannot at this point
confirm that these patterns are generic features and that they will hold for
other extensions. Again, we recall that the results of the order-eight model
should be taken with care, since the solutions are not global minima. The
global minima which would lead to negative energy and a zero-size soliton is
prevented by a finite lattice spacing. But the order-eight model and its
solutions should be considered here as good approximations for more elaborate
models such as (\ref{allorder}) as can be seen from table \ref{Sk_vs_O6_O8}.

\begin{table}[ptb]
\begin{center}%
\begin{tabular}
[c]{|l|c|c|c|c|c|c|c|c|c|}\hline
& \multicolumn{3}{|c|}{$\chi_{S}$ (Skyrme)} & \multicolumn{2}{|c|}{$\chi_{O6}$
(order 6)} & \multicolumn{2}{|c|}{$\chi_{O8}$ (order 8)} &
\multicolumn{2}{|c|}{$\chi_{R}$ (all orders)}\\
& \multicolumn{3}{|c|}{} & \multicolumn{2}{|c|}{} & \multicolumn{2}{|c|}{} &
\multicolumn{2}{|c|}{}\\
& RM & 3D & 3D \cite{Hale2000} & RM & 3D & RM & 3D & RM & 3D\\\hline
$B_{1}$ & 0.9999 & 0.9974 & 1.0015 & 0.9999 & 0.9984 & 1.0000 & 0.9995 &
0.9999 & 0.9994\\
$E_{1}$ & 103.1306 & 103.091 & 104.45 & 129.9757 & 130.828 & 129.2987 &
130.242 & 129.3145 & 130.156\\\hline
$B_{2}$ & 1.9999 & 1.9981 & 2.0030 & 2.0000 & 1.9995 & 2.0000 & 1.9998 & --- &
1.9996\\
$E_{2}$ & 202.3558 & 198.139 & 199.17 & 253.2955 & 245.568 & 252.0314 &
244.542 & --- & 244.609\\
$E_{2}/E_{1}$ & 1.9621 & 1.9219 & 1.9068 & 1.9488 & 1.8770 & 1.9492 & 1.8776 &
--- & 1.8793\\\hline
$B_{3}$ & 3.0000 & 2.9981 & 3.0042 & 2.9999 & 2.9997 & 3.0000 & 2.9999 & --- &
2.9998\\
$E_{3}$ & 297.5660 & 288.641 & 291.11 & 369.3230 & 354.405 & 367.7119 &
352.940 & --- & 352.906\\
$E_{3}/E_{1}$ & 2.8853 & 2.7998 & 2.7870 & 2.8415 & 2.7089 & 2.8439 & 2.7099 &
--- & 2.7114\\\hline
$B_{4}$ & 4.0001 & 3.9989 & 4.0048 & 3.9999 & 3.9997 & 4.0000 & 3.9997 & --- &
3.9997\\
$E_{4}$ & 380.7257 & 377.281 & 375.50 & 467.3910 & 455.805 & 465.4414 &
454.205 & --- & 454.190\\
$E_{4}/E_{1}$ & 3.6917 & 3.6597 & 3.5950 & 3.5960 & 3.4840 & 3.5997 & 3.4874 &
--- & 3.4895\\\hline
\end{tabular}
\end{center}
\caption{Results of SA in one dimension using the rational map ansatz (RM) and
in three dimensions (3D) for the Skyrme model and extensions defined in
(\ref{order246}, \ref{allorder}). For the Skyrme model the 3D results are
compared with those of ref. \cite{Hale2000}. $B_{N}$ and $E_{N}$ are the
computed baryonic number and energy for the soliton configuration with winding
number $N$ respectively. The unit of energy is $F_{\pi}/(2\sqrt{2}e)$.}%
\label{Sk_vs_O6_O8}%
\end{table}

\begin{figure}[ptb]
\begin{center}
\includegraphics[scale=0.4, angle=270]{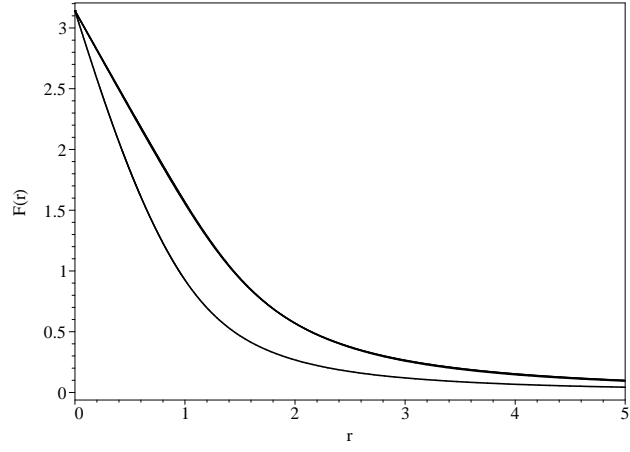}
\end{center}
\caption{Chiral angles obtained with one dimensional SA for the Skyrme model
and extensions of order six and order eight in the topological sector $B=1$.
The lower curve corresponds to the $F(r)$ of the Skyrme model. The two
higher-order extensions we have studied lead to superposed chiral angles
decreasing slower that the one of the Skyrme model.}%
\label{Plot_Fr}%
\end{figure}

\begin{figure}[ptb]
\begin{center}
$\qquad\qquad B=1 \qquad\qquad\qquad\qquad B=2 \qquad\qquad\qquad\qquad\quad
B=3 \qquad\qquad\qquad\qquad B=4$\newline%
\includegraphics[scale=0.15, angle=270]{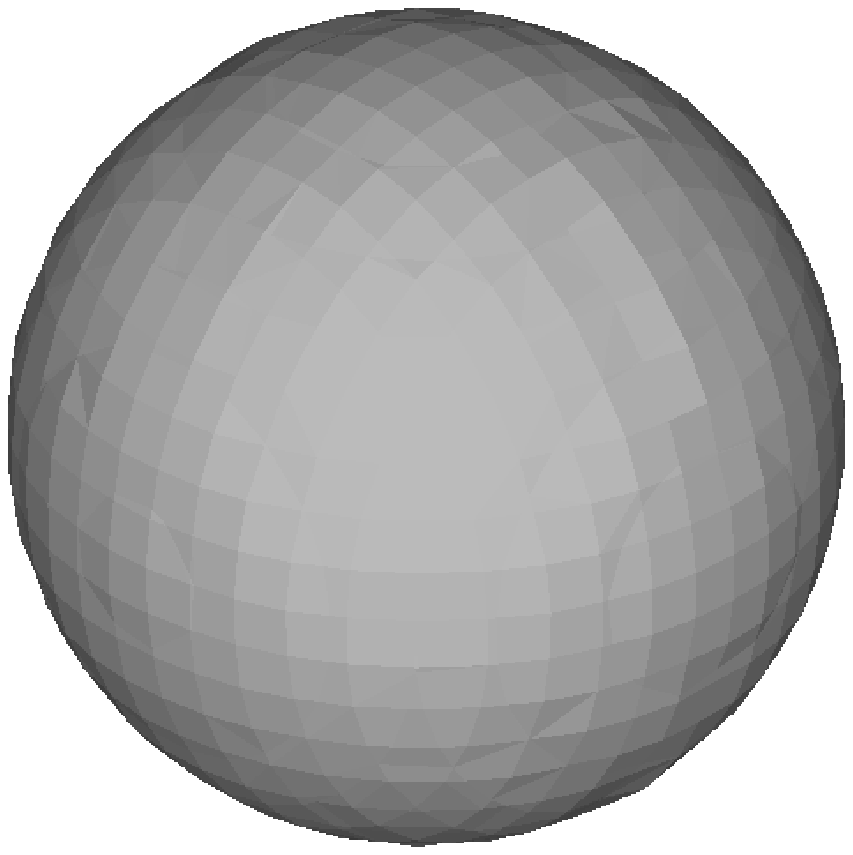}
\includegraphics[scale=0.15, angle=270]{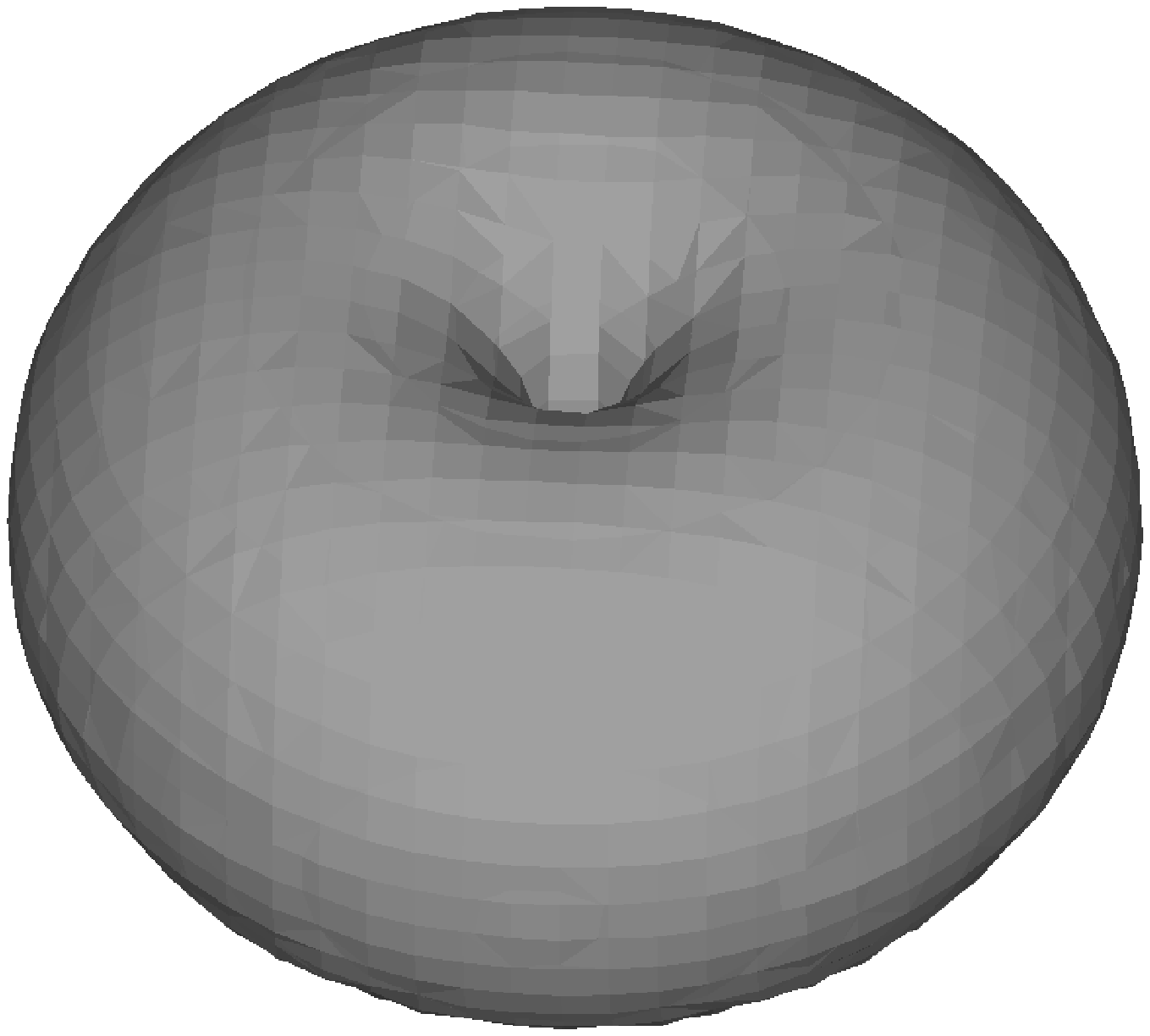}
\includegraphics[scale=0.15, angle=270]{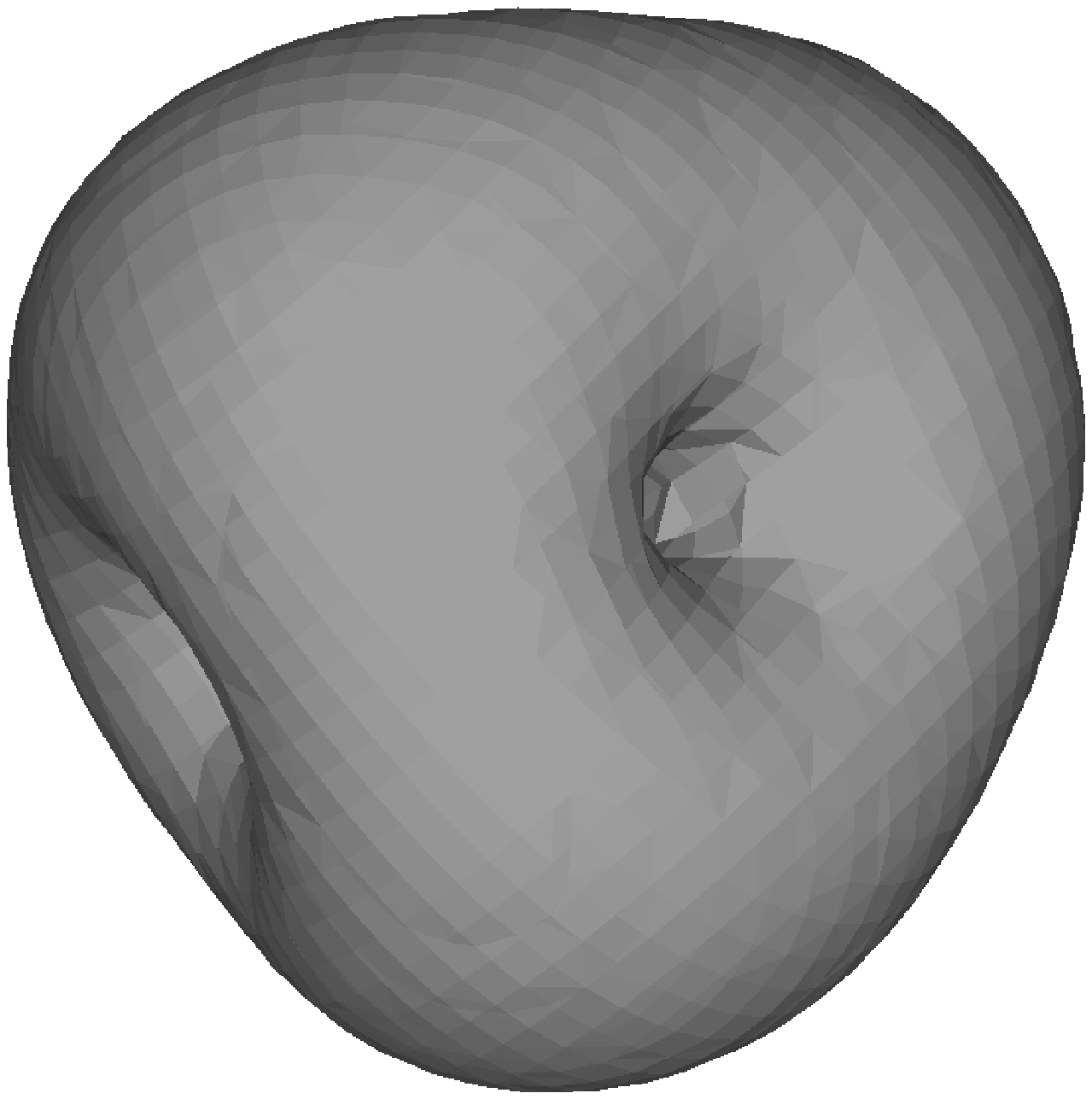}
\includegraphics[scale=0.15, angle=270]{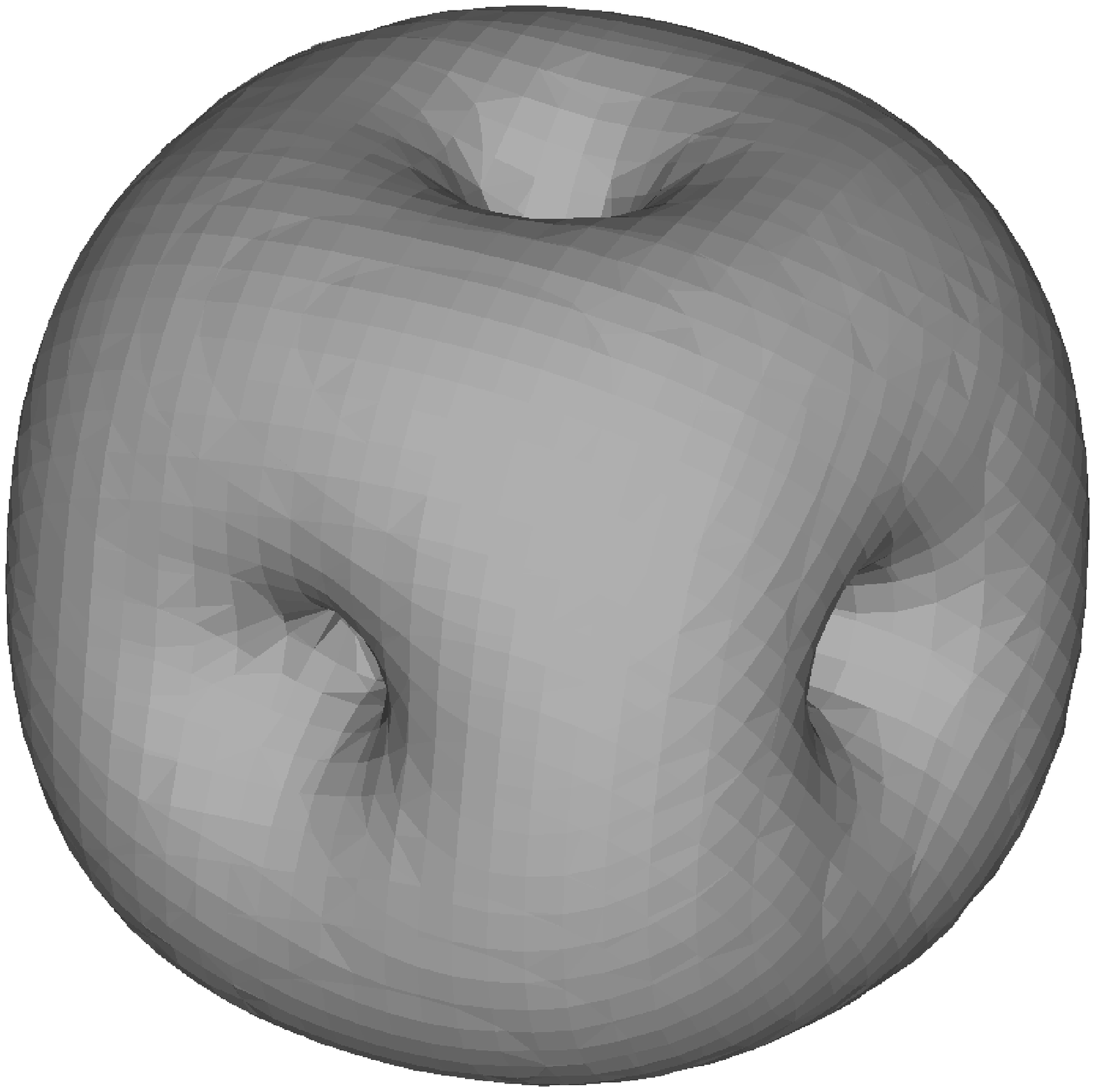}\\[0.4cm]%
\includegraphics[scale=0.15, angle=270]{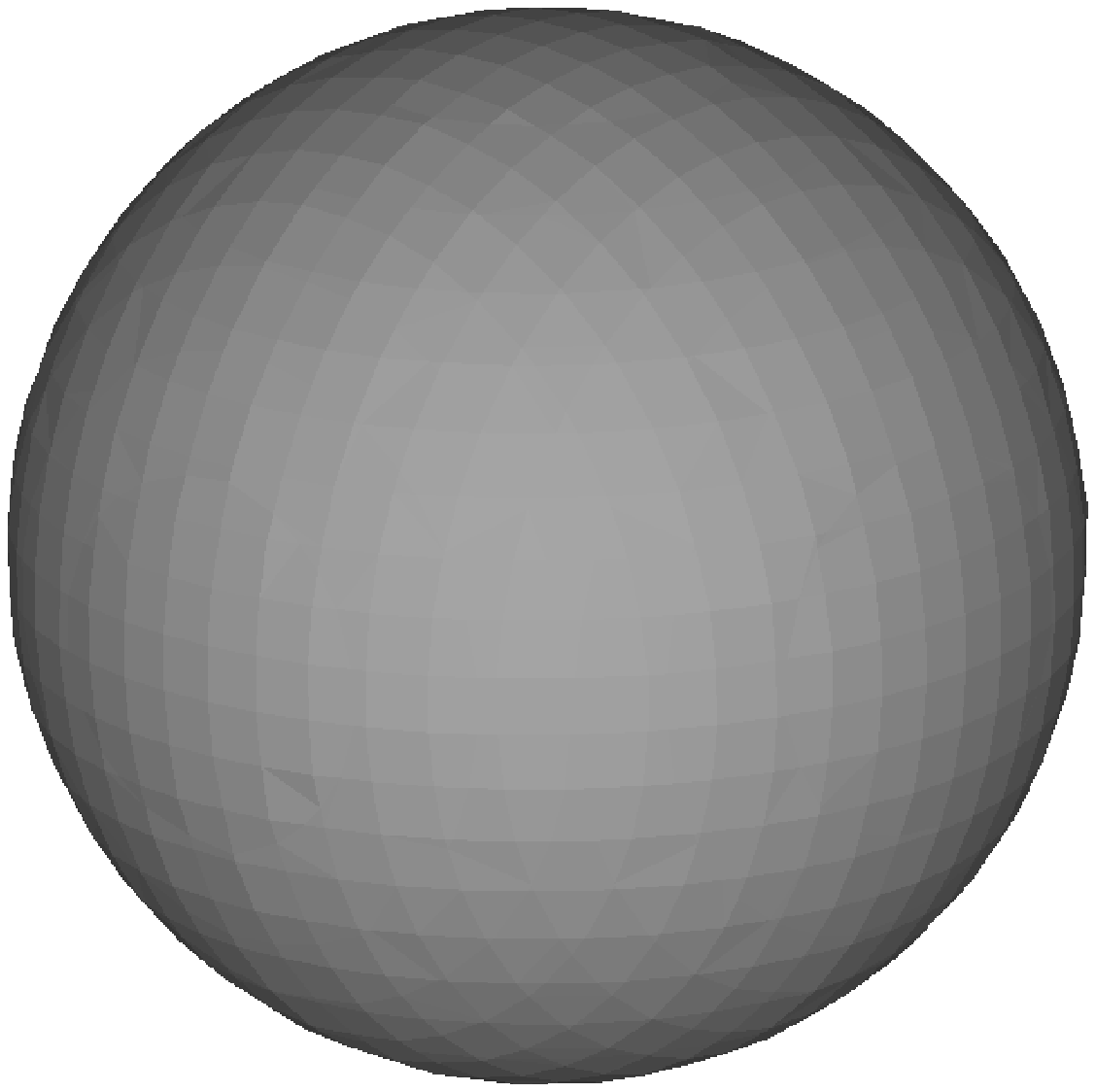}
\includegraphics[scale=0.15, angle=270]{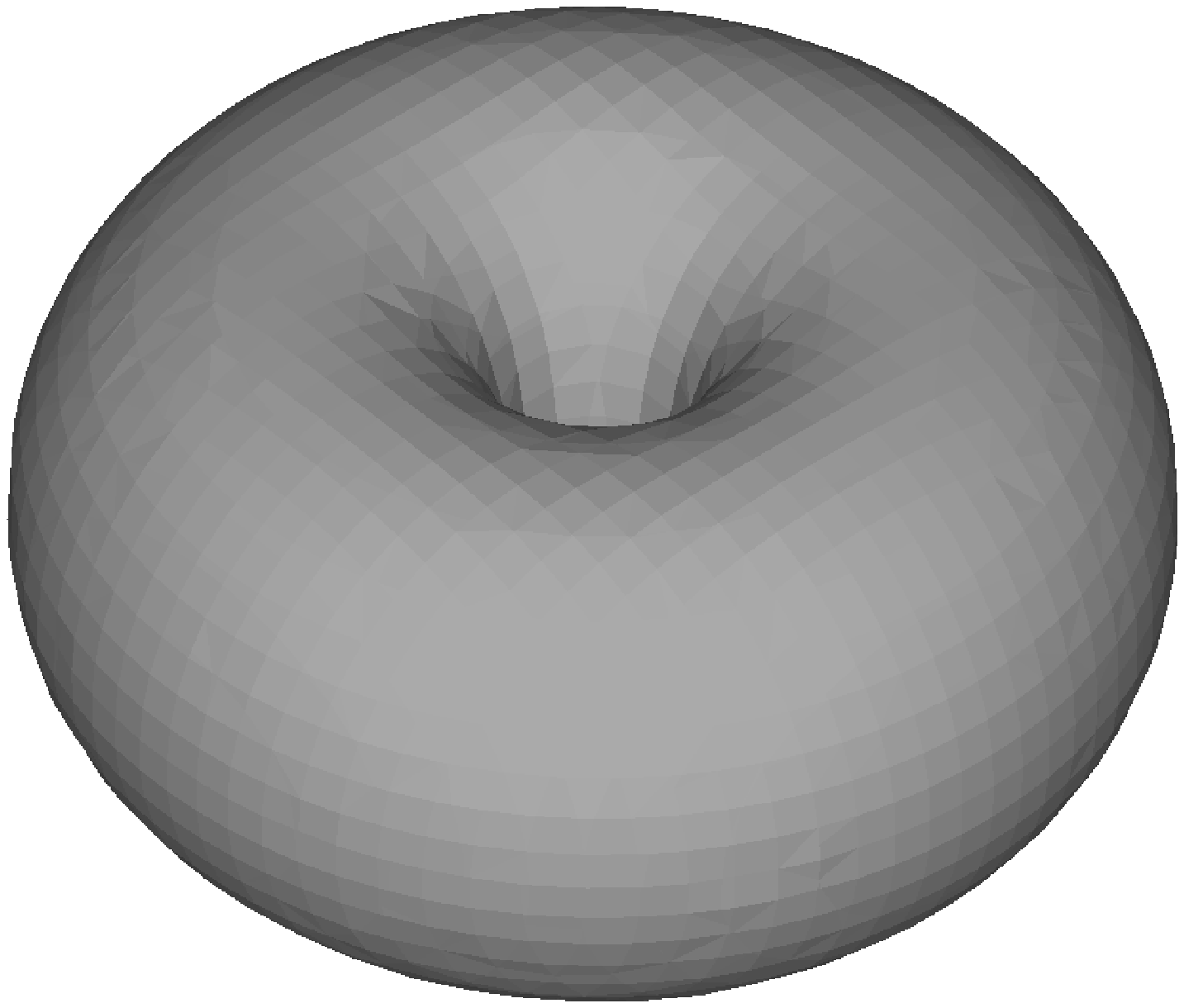}
\includegraphics[scale=0.15, angle=270]{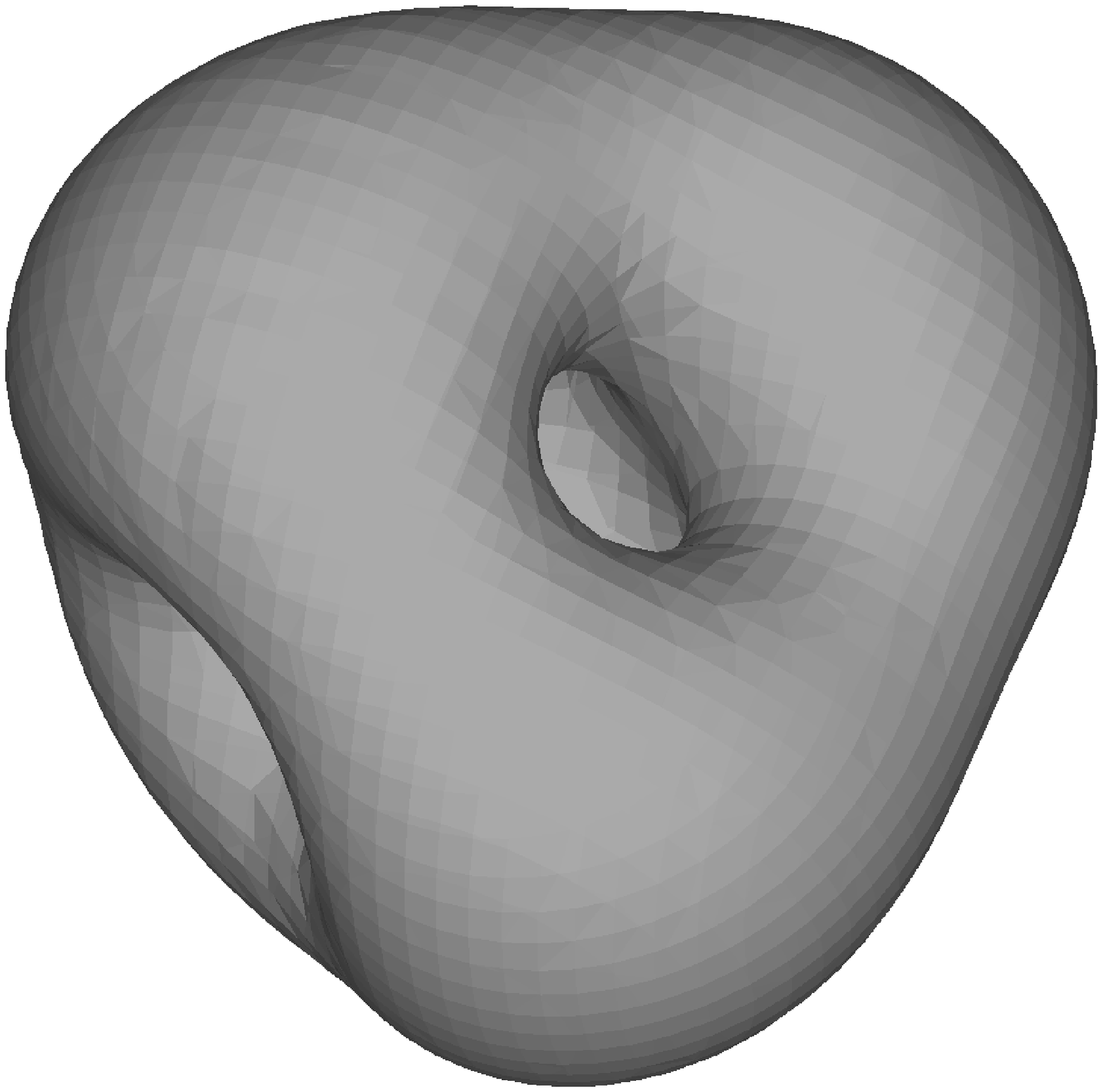}
\includegraphics[scale=0.15, angle=270]{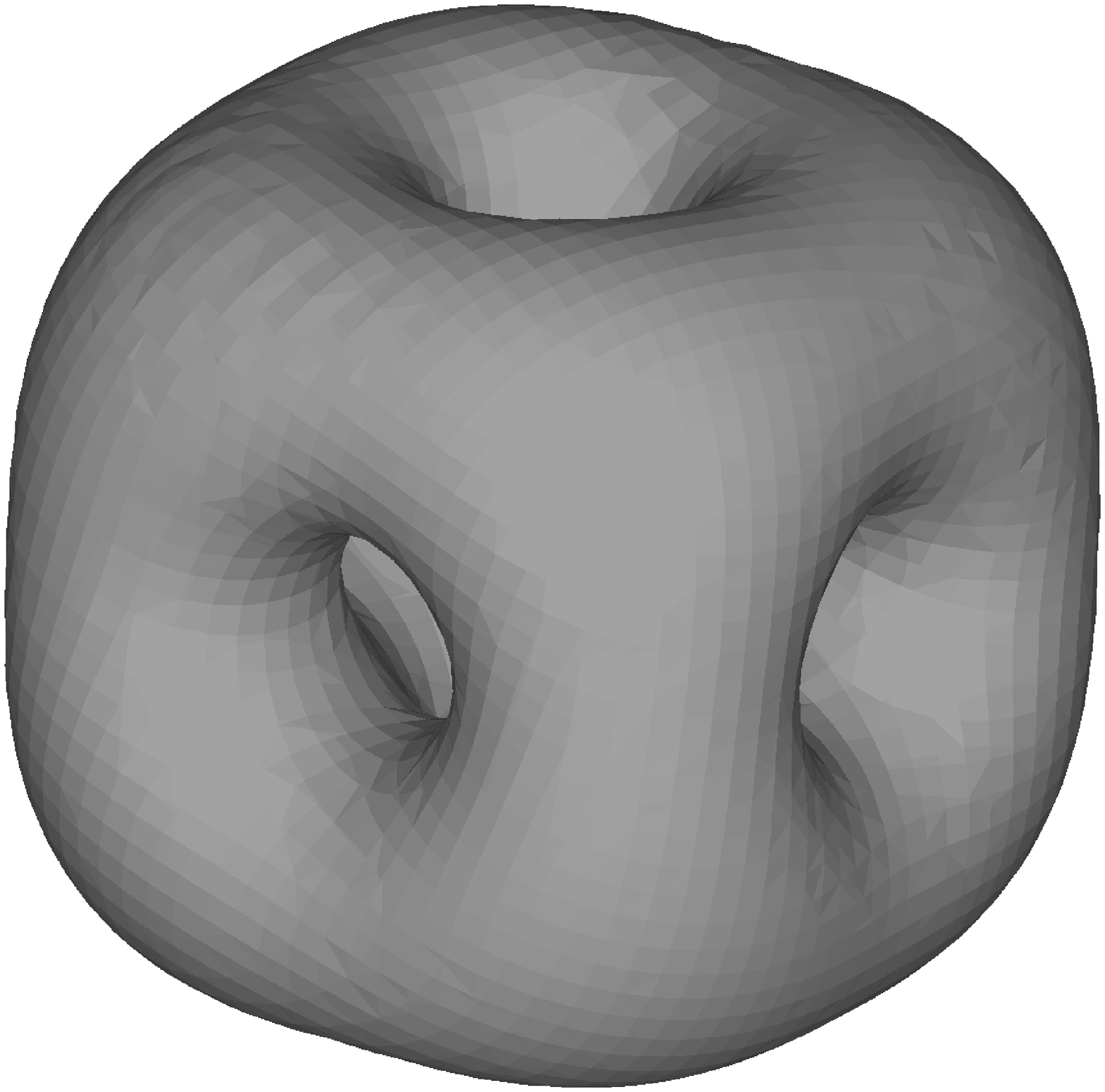}\\[0.4cm]%
\includegraphics[scale=0.15, angle=270]{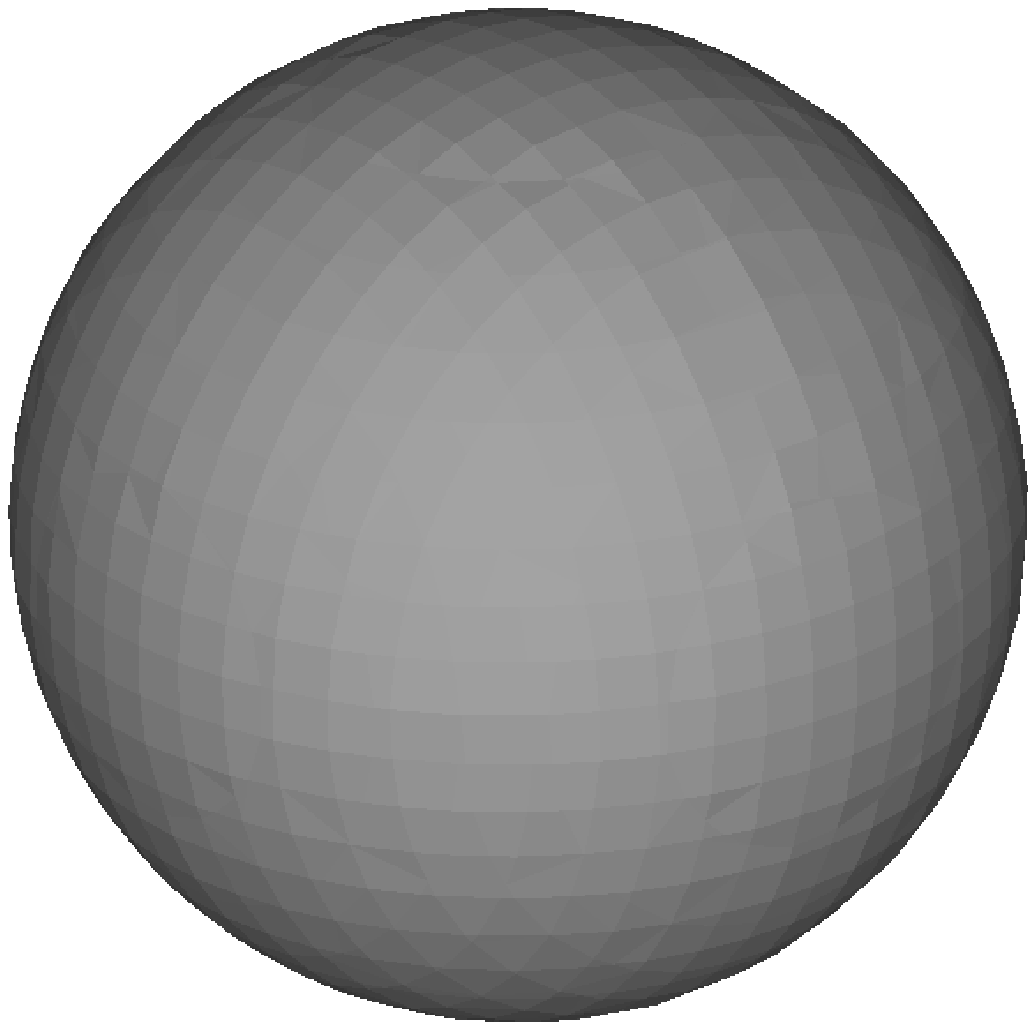}
\includegraphics[scale=0.15, angle=270]{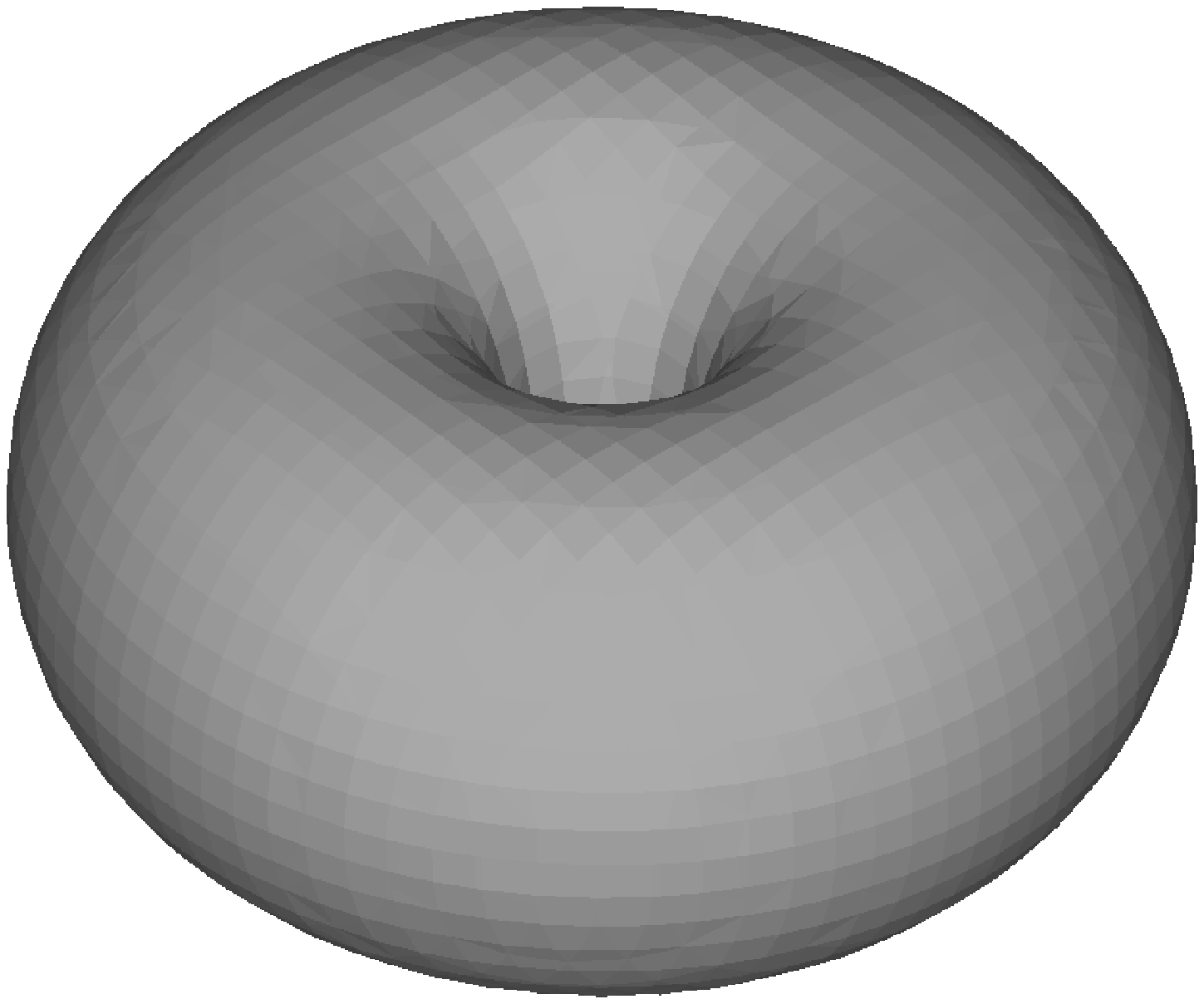}
\includegraphics[scale=0.15, angle=270]{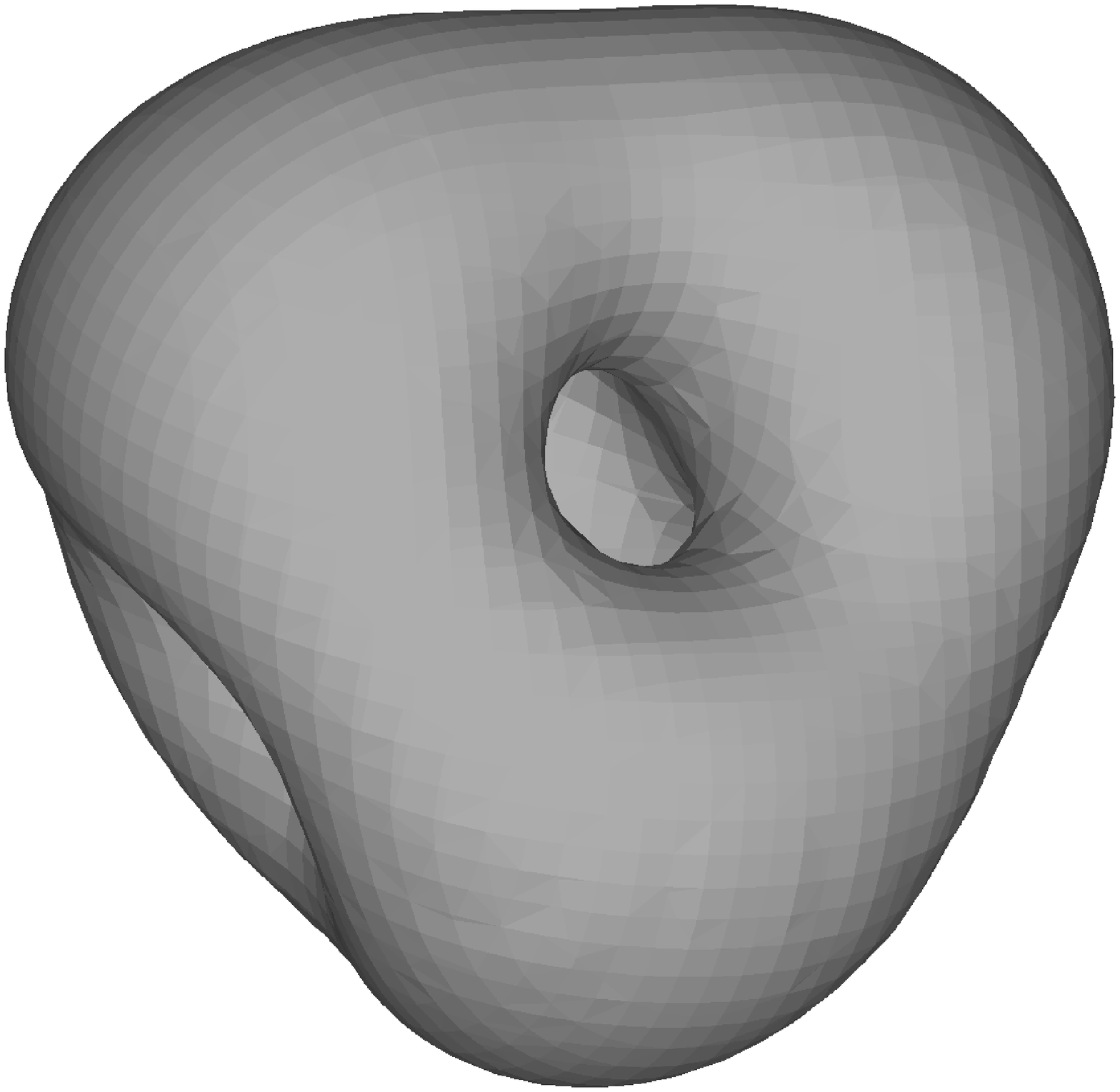}
\includegraphics[scale=0.15, angle=270]{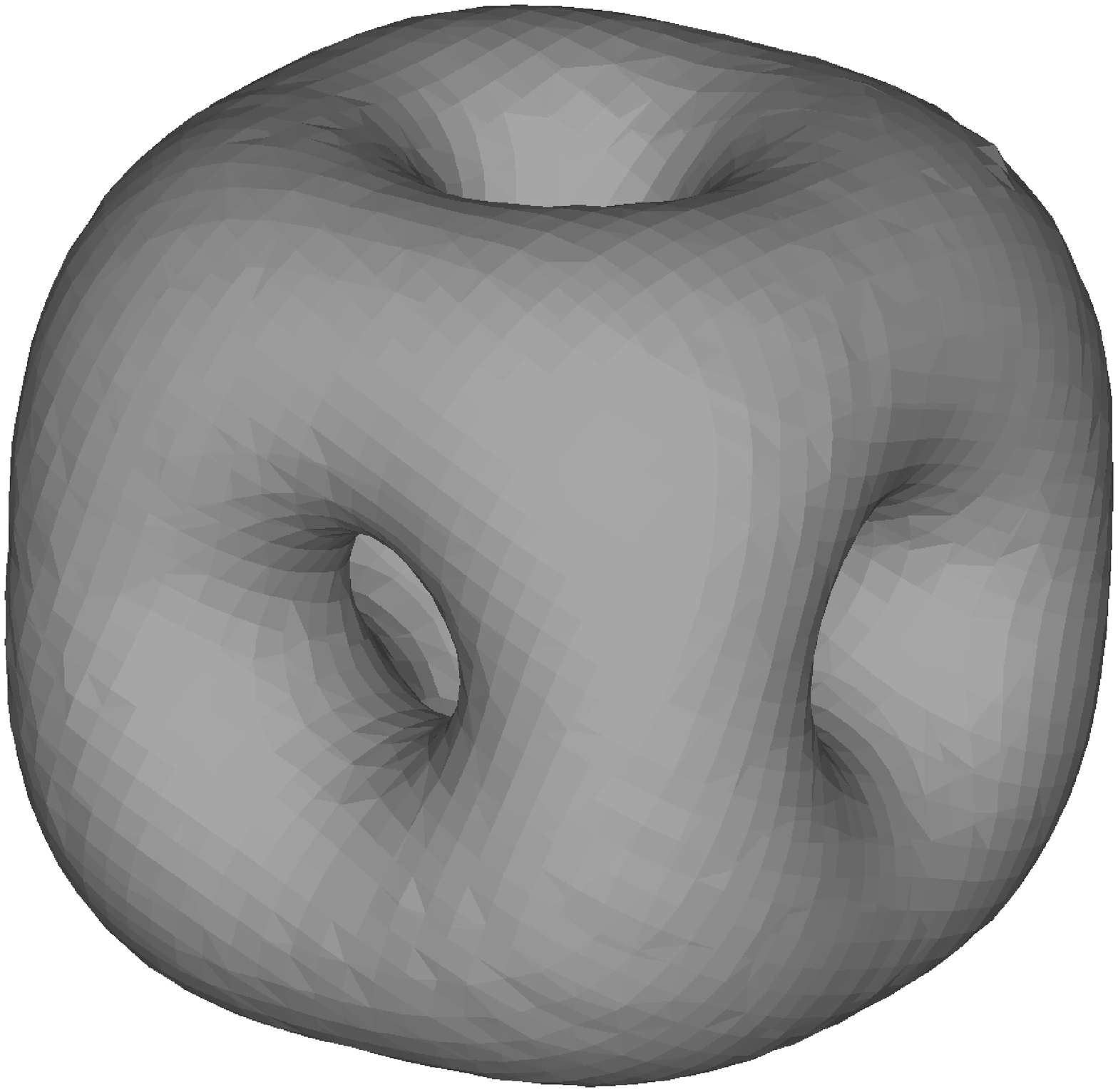}
\end{center}
\caption{Plots of constant baryonic density ($0.0001$) obtained using three
dimensional SA. From top to bottom are shown the results of the Skyrme model,
the order-six extension and the order-eight extension.}%
\label{W_const_figs}%
\end{figure}

The increase in the static energy of generalized skyrmions relative to those
of the Skyrme model reflects itself in the structure of such solutions. The
plots of constant baryonic density illustrated on figure \ref{W_const_figs}
clearly show that extended skyrmions occupy a greater volume. This fact can
also be supported by the rational maps approximation with one dimensional SA
calculations (see the chiral angles in figure \ref{Plot_Fr}). The plots of
figure \ref{W_const_figs} also strongly suggest that the symmetries of
extended skyrmions are the same as those of the Skyrme model.

\begin{table}[ptb]
\begin{center}%
\begin{tabular}
[c]{|l|c|c|c|c|}\hline
& Skyrme & Order 6 & Order 8 & Rational Model\\\hline
$B=1$ & $0.04$ & $0.65$ & $0.72$ & $0.65$\\
$B=2$ & $2.13$ & $3.15$ & $3.06$ & ---\\
$B=3$ & $3.09$ & $4.21$ & $4.19$ & ---\\
$B=4$ & $0.91$ & $2.54$ & $2.47$ & ---\\\hline
\end{tabular}
\end{center}
\caption{Differences in percent ($\%$) between the static energies of the
solutions obtained using the rational maps ansatz and the ones obtained using
three dimensional SA.}%
\label{Approx_percent}%
\end{table}

Let us now make quantitative remarks regarding the reliability of the rational
maps ansatz. According to the conclusions of \cite{Houghton1998}, the rational
maps ansatz should lead to computed static energies of $B>1$ for the original
Skyrme model that are larger by a few percent compared to the exact solutions.
Indeed, using the data of table \ref{Sk_vs_O6_O8}, we see that this is true
for the Skyrme model, as well as for the two extensions we have considered.
The differences between the two approaches are reported in table
\ref{Approx_percent}. It seems that adding new terms to the Skyrme model
slightly compromises the reliability of the rational maps ansatz. However, the
discrepancies are quite moderate and the rational maps approximation remain
valid for extensions of the Skyrme model. Therefore, both baryonic density
plots and the comparison with rational map ansatz indicate that the solutions
of the Skyrme model and those of the generalized models we have studied are
characterized by a similar angular distribution, i.e. the same symmetries.

Even if SA leads to satisfying soliton solutions, some numerical features that
are inherent to the simulation occur and it is important to review their
effects on our results. First of all, we note that despite what we expected,
the static energies of the $B=1$ solitons in the 1D and 3D schemes do not
exactly coincide. This is mostly due to a larger spacing of the 3D lattice,
which must be imposed in regard to computational time and memory. Moreover,
the finite volume of the 3D (periodic) lattice induces an error estimated at
$1\%$ \ \cite{Hale2000}, because the soliton interacts with itself over the
boundaries. Another source of error is due the non logarithmic nature of our
cooling schedule which amounts to about $0.1\%$. Finally, an error of the
order of $0.3\%$ is expected to arise as a result of the way we evaluate the
derivatives on the finite spacing lattice.

For the sake of completeness, we now compare our 3D results (table
\ref{Sk_vs_O6_O8}) with existing calculations. For the Skyrme model, previous
calculations that were performed using several approaches, for example an
axially symmetric ansatz \cite{Bratten1988,Kopeliovich1987}, a relaxation
method \cite{Bratten1990,Battye19971, Battye19972} and more recently SA
\cite{Hale2000} are in accord with the results presented in this work.
Kopeliovich and Stern \cite{Kopeliovich1987} and Floratos et al.
\cite{Floratos2001} have also analyzed some Skyrme model extensions up to
order six but the choice of weights $h_{3}$ and $h_{4}$ are different so our
model cannot be compared directly. However, their solutions exhibit the same
symmetries as in figure \ref{W_const_figs}. The remaining results of table
\ref{Sk_vs_O6_O8} are completely new.

Even if the results that we have obtained so far support the rational map
conjecture proposed in this work, its general character needs to be
investigated in greater detail. There are indeed indications that a simple
modification to the Lagrangian may change the form of the soliton, e.g. adding
a mass term \cite{Battye2004}. As a starting point, one could analyze other
extended models obeying the positivity constraints presented in section
\ref{sec:Positivity}. The SA algorithm is particurlarly well suited to this
kind of problem since there is no need to handle complex differential
equations. It could also be interesting to add a pion mass term or to study
models having different structures, for example a contribution coming from the
rotational energy of the skyrmion. In fact, the versatility of the SA
algorithm makes it possible to study a large variety of models. Some
investigations along these paths are already under way.

\end{document}